%% file: astroph_cal.tex
\documentstyle[aps,prl,preprint,floats,epsfig]{revtex}
\newif\iftightenlines\tightenlinesfalse
\tightenlines\tightenlinestrue

\newcommand{\be}{\begin{equation}} \newcommand{\ee}{\end{equation}}
\newcommand{\ba}{\begin{eqnarray}} \newcommand{\ea}{\end{eqnarray}}
\newcommand{\nn}{\nonumber}

\begin{document}

\tighten

\title{Performance and Simulation of the RICE detector}

\input riceauthor

\maketitle

\begin{abstract}
The RICE experiment (Radio Ice Cherenkov Experiment) at the South Pole,
co-deployed with the AMANDA experiment,
seeks to detect ultra-high energy
(UHE) electron neutrinos interacting in cold
polar ice. Such interactions produce electromagnetic showers,
which emit
radio-frequency Cherenkov radiation.
We describe the experimental apparatus
and the procedures used to measure the
neutrino flux.
\end{abstract}

\section{Ultra-High Energy Neutrino Astronomy: Introduction}
Detection of 
ultra-high energy ($E_\nu > 10^{15}$eV) neutrinos represents a 
unique opportunity to probe the distant universe. High-energy protons and
photons from distant sources
are likely to interact with the cosmic microwave background (CMB);
protons, being charged, have their trajectories bent in galactic and 
intergalactic magnetic fields. Neutrinos are
inert to CMB photons and point 
directly back to their source, giving 
essential information on those sources.
In the realm of particle physics, detection of UHE neutrinos 
from cosmological distances, if 
accompanied by flavor identification, may permit measurement 
of neutrino oscillation parameters 
over an unprecedented range of $\Delta m^2$. 
It has been suggested that, with a sensitive enough array,
tau neutrinos may be identified by two-step 
``double-bang"\cite{doublebang} processes where a $\tau$ lepton is 
created and subsequently decays.
Since neutrino absorption in the earth
depends on the chord length through the
earth,\footnote{The Earth is opaque to
$>$PeV neutrinos with Standard Model
cross-sections and with zenith angles approaching 180 degrees.
Largely because of earth shadowing, the RICE array is most
sensitive to neutrinos incident at
zenith angles between 60 and 120 degrees. Conversely, the angular
distribution is sensitive to the cross-section and may allow
checks of the Standard Model.}
the angular distribution of detected neutrino events
could be used to verify predictions for weak 
cross-sections at energies unattainable by any man-made accelerator.
Alternately, if the high energy weak cross-sections are known,
they can be used to test Earth composition models
along an arbitrary chord (`neutrino tomography')\cite{tomography}.

\subsection{Current Experimental Efforts}
Several recent projects 
(including AMANDA\cite{AMANDA99}, 
NESTOR\cite{NESTOR}, Lake Baikal\cite{BAIKAL99}, 
ANTARES\cite{Antares00}) 
are optimized for detection of very
high energy ($10^{12-15}$ eV) cosmic ray 
muon neutrinos.
Sensitivities to higher energies, as well as
electromagnetic cascades, have also been shown to be 
substantial in such experiments\cite{TWR,ICRC-papers}.
These instruments are based on photomultiplier tube 
detection of the optical Cherenkov cone from muons
produced in muon
neutrino charged current interactions. At high energies,
muons have ranges of
order 1 km and follow approximately straight trajectories
(smeared by multiple scattering), punctuated by
catastrophic bremstrahlung every 0.1--1 km or so, 
in which $\sim$10\% of the
muon's energy is lost to a photon.
RICE employs radio detection, which is believed to be
the most efficient 
detection mechanism at energies 
of $10^{15}$ eV and beyond\cite{Buford}.

\begin{figure}[htpb]
\begin{picture}(200,250)
\includegraphics{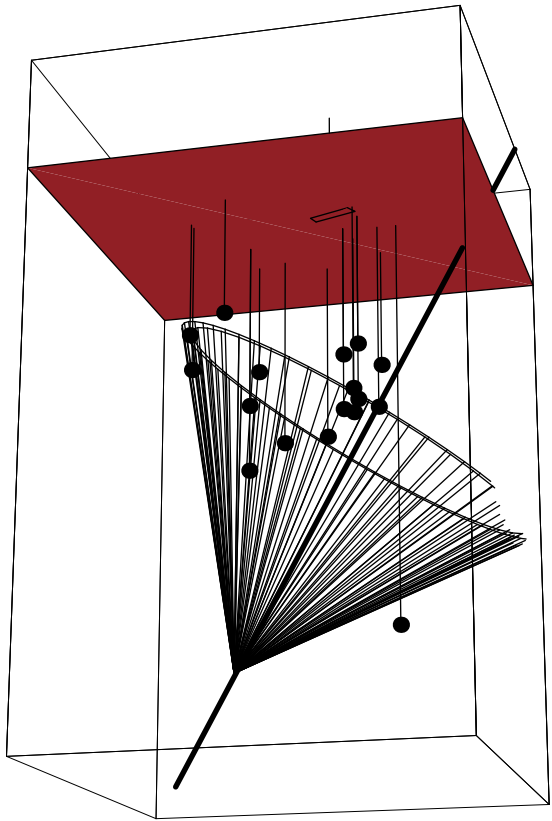}
\end{picture} 
\caption{Simulated RICE event. The actual detector
geometry is shown, to scale.}\label{fig:evdisp}
\end{figure}
The RICE concept is illustrated in 
Figure \ref{fig:evdisp}, depicting a 
Cherenkov cone fit to a set of ``struck'' dipole receivers in
a simulation event, along with the
extracted neutrino direction.
Receiver locations are drawn to scale in the Figure.
In the actual array geometry, dipole receivers are
spread over a 200 m $\times$ 200 m $\times$ 200 m cube beneath and
around the Martin A. Pomerantz Observatory (MAPO), approximately 1 km.
from the geographic South Pole.

\subsection{Radio Detection}
We have initiated a pilot program
based on radiowave receiver
technology in order to extend 
electron neutrino detection up to the PeV
energy scale.
The detector is intended to have sensitivity in the energy
range $E_{\nu_e}\sim 10^{15}-10^{18}$ eV.
RICE (``Radio Ice Cherenkov Experiment'')
should ultimately be capable of observing the sky with
angular resolution of $\sim$10-50 mrad.

Coherent radio Cherenkov emission is an
efficient method for
detecting high energy particles.  The history of the effect goes back to
Jelley, who first considered whether cosmic ray air showers might produce a
radio signal\cite{Allan}.  Askaryan\cite{Askaryan} subsequently
predicted a net charge imbalance in air showers, and coherent radio power
proportional to the energy of the shower squared.  Substantial radio emission
from atmospheric electromagnetic cascades was observed more than 30 years
ago\cite{Allan,Jelley}.  Progress in ultra-high energy air showers has
sparked renewed interest, and new observations of radio pulses have been
reported recently\cite{Rosner}, suggesting a 
possible radio component to the
Auger detector\cite{Auger00}.
A recent international meeting highlights current progress\cite{RADHEP2000}.

\begin{figure}[htpb]
\centerline{\includegraphics[width=10cm]{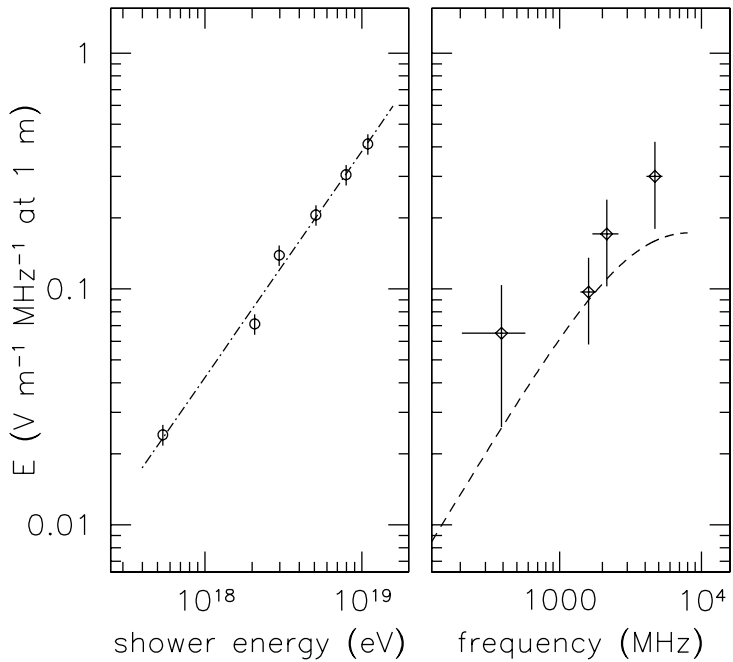}}
\caption{Results of testbeam experiment[14], showing
signal strength and expected dependence on total beam current
(circles,
left) and frequency (diamonds,
right). Figure reprinted courtesy of D. Saltzberg and
P. Gorham.}
\label{slac_fig4.ps}
\end{figure}

This effect has recently been observed in a test-beam experiment at
SLAC\cite{testbeam}. A beam of electrons, of known amperage, was fired into a 
sand target and the radiation resulting from the impact measured in
the radio regime. As Figure \ref{slac_fig4.ps} illustrates, the
measured signal strength (diamonds) displays the expected  
dependence on both total beam current (left, solid curve) and 
frequency (right, dashed curve).

\subsection{Initiation of the RICE Experiment}
The RICE experiment was initiated in October, 1995 when 
the AMANDA collaboration graciously consented
to co-deployment of two shallow
radio receivers (``Rx'') in the first 
holes being drilled for AMANDA-B. 
Following deployment, a surface transmitter (``Tx'')
was used to verify that signals could be detected under-ice
with better than 10 ns timing precision. 
However, cross-talk and 
amplifier oscillation
problems precluded use of those receivers for
science.
The first three dedicated RICE
receivers were deployed in 1996-97, along with one underice 
transmitter. The extraordinary RF clarity of South Polar ice was amply
illustrated by the evident brightness of the AMANDA photomultiplier
tubes 2 km. below the RICE receivers (and close to
the null in the dipole antenna's reception pattern). 
A Fourier analysis of the RF
transients
produced by AMANDA photo-tubes in
laboratory conditions indicated that PMT background power
dominated at frequencies below 100 MHz, motivating use of 
high-pass filters in subsequent Rx deployments. 
Antenna deployments
followed in
1997-98 (three more receivers and two more transmitters deployed in
AMANDA-B holes), 1998-99 (six receivers deployed in 5''-diameter
dedicated RICE holes, bored using a mechanical, rather than a hot-water
drill), and 1999-2000 (six receivers, and
one transmitter deployed in AMANDA-B holes).

\section{Experimental Apparatus}
\label{intro.sec}
The RICE
experiment presently consists of an 18-channel array of radio 
receivers (``Rx''),
scattered within a 200 m$\times$200 m$\times$200 m cube, at 100-300 m depths.
Each receiver contains a half-wave dipole antenna, offering 
good reception over the range 0.2--1 GHz.
Twelve receivers are buried in the 
boreholes drilled for the AMANDA photomultiplier tube
deployment during the 1996-97, 97-98, and 99-00 
austral summers. Six receivers
are located in dedicated RICE holes; 
four such holes were drilled with
a 5-inch diameter mechanical hole-borer in 1998-99. 
The signal from each antenna is 
immediately boosted by a 36-dB in-ice amplifier, then
carried by  $\sim$300 m
coaxial cable to the surface observatory, where the signal is
filtered (suppressing noise below 200 MHz due to both AMANDA
photo-tubes, as well as continuous wave backgrounds from
South Pole station at 149 MHz), re-amplified
(either 52- or 60-dB gain), and split into two copies.
One copy is fed into a
CAMAC crate from which,
after initial discrimination (using a LeCroy 3412 discriminator),
the signal is routed into a NIM crate where the trigger logic
resides. The other copy of the analog signal from the antennas is input
to one channel of a digital oscilloscope, where waveform information
is recorded.
Also deployed are three large TEM 
surface horn antennas which are used as a veto of surface-generated noise.


\subsection{Detector Array Status}
\begin{figure}[htpb]
\includegraphics[width=16.5cm,angle=0]{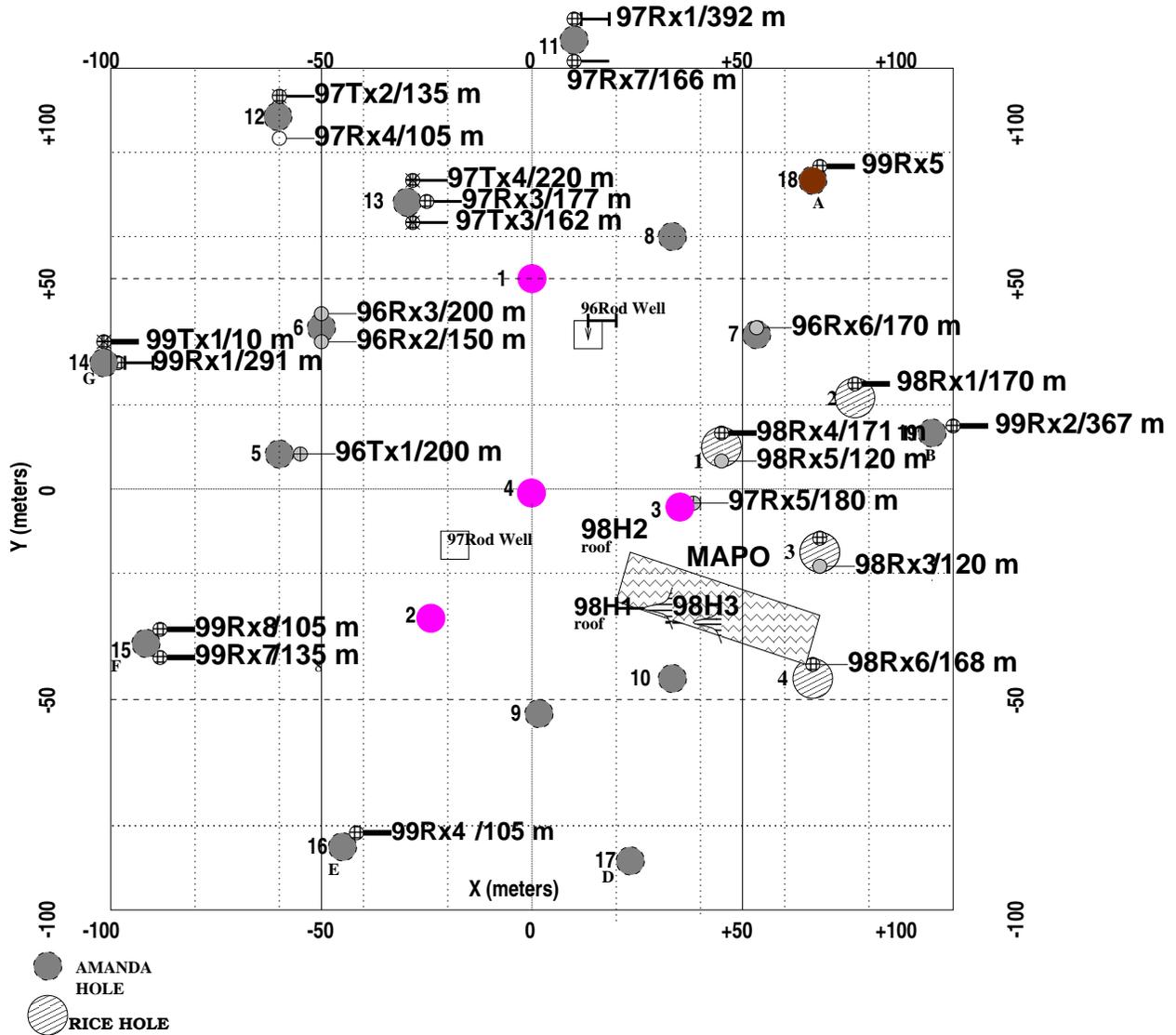}
\caption{Present geometry of the RICE array, relative to AMANDA
hole 4. ``Tx'' designate transmitters; ``Rx''
designate receivers. Depths of receivers, as well as 
relative cable diameter
(indicated by the thickness of horizontal lines before the 
antenna identifier), 
is also indicated.}
\label{fig:rice_rxtx}
\end{figure}

The status of the current array
deployment is summarized in Figure \ref{fig:rice_rxtx}. 
Indicated in the Figure are the AMANDA holes (1-19, drilled using
the hot-water technique) containing RICE receivers, and
also the four holes drilled specifically for RICE in 1998-99 
containing RICE-only equipment.
All channels are indicated by
a 5-character alphanumeric mnemonic corresponding to the year of deployment,
the type of dipole antenna deployed
(transmitter ``Tx'' or receiver ``Rx''; these only
differ by the installation of a receiver
amplifier in the latter modules), and a numerical identifier.
Also indicated
in the Figure is the MAPO building which houses 
hardware for several experiments, including
the RICE and AMANDA surface
electronics. The AMANDA array is located approximately 600 m (AMANDA-A) to
1400 m (AMANDA-B) below the RICE array in the ice; the 
South Pole Air Shower Experiment
(SPASE) is
located on the surface at $x\sim-400$m in the Figure.
The coordinate system conforms to the convention used by the
AMANDA experiment; grid North coincides with the +y-direction in the Figure.

\subsection{Event Trigger and Vetoes}
The main RICE trigger condition requires that four or 
more channels have their output voltage exceed a common adjustable 
threshold within a pre-set time window (currently $\Delta t=1.2\mu$s,
set by the radio wave transit time across the array). The 
threshold is adjusted so that thermal fluctuations, or other backgrounds, 
do not cause excessive triggers. The time window allows for coincidences 
across the full array, independent of event geometry. 
A valid event trigger is also defined when,
in a time window of 1.2 $\mu$s,
i) $\ge$ one underice antenna registers a signal
above threshold in coincidence with a 30-fold PMT AMANDA-B trigger, or:
ii) 
$\ge$ one underice antenna registers a signal above threshold in coincidence
with a high-multiplicity
SPASE event.

Additionally, there 
are two ways that surface-generated background transient
events can be vetoed -- either: 
a) one of the surface horn antennas registers
a signal above threshold,
in which case
data-taking is inhibited for the subsequent 4 $\mu$sec,\footnote{The
raw rate for this veto is of order 10 Hz, so this represents an
insignificant loss of data.}
or 
b) the timing sequence of hits in the underice antennas is
determined to be consistent (in software) 
with the sequence expected from 
surface-generated backgrounds.\footnote{In 
the event of an AMANDA or SPASE coincidence trigger,
the surface veto is disabled in order to preserve possible 
air shower events.} 
This ``surface-veto'' algorithm uses the time
differences between TDC hits recorded on-line to make a fast ($<$10 msec)
classification of the event as surface/non-surface in 
origin.\footnote{This functionality will reside in a
dedicated CAMAC board beginning in 2002.} One of
every 10000 events classified as ``surface'' (i.e., having
a sequence of antenna hits consistent with a $Z>$0 vertex)
are retained to ensure that this veto is functioning properly.

\begin{figure}[htpb]
\includegraphics[width=5in,angle=-90]{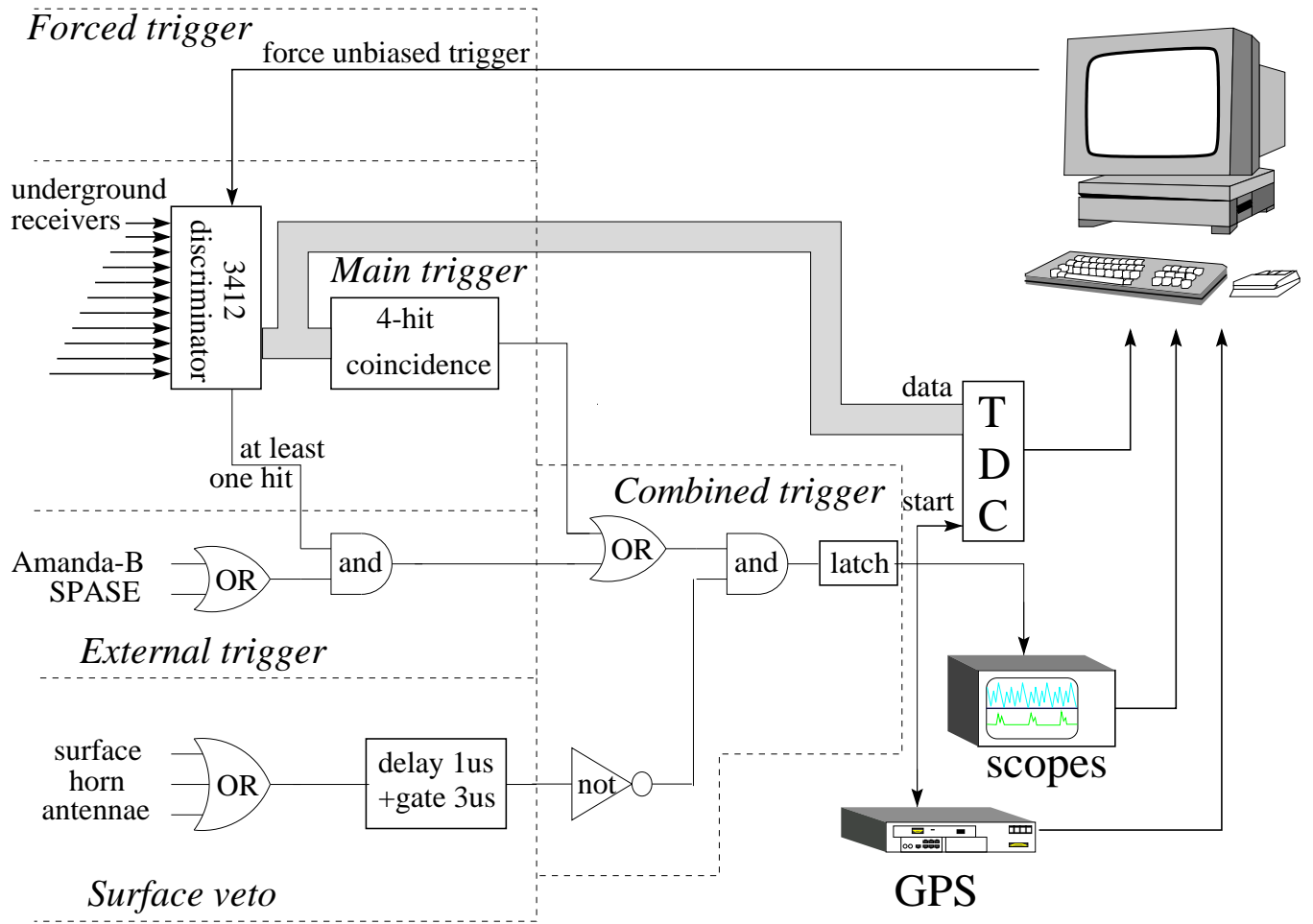}
\vspace{0.4cm}
\caption{Functional
schematic of RICE event trigger.}
\label{fig:function.ps}
\end{figure}

If any of the above trigger
criteria are satisfied and there is no veto 
signal,
the time of each hit above
threshold (as recorded by a LeCroy 3377 TDC), and also an 8.192 $\mu$s buffer
of data stored in an HP54542
digital oscilloscope at 1 GSa/s (for each channel) is written to disk.
Each event is also
given a GPS time stamp for synchronization with other South Pole
(and more global) experiments.
To monitor
the stability of the amplifiers as well as changing background
conditions, waveform measurements are taken every ten minutes,
independent of other event triggers (so-called
``unbiased'' triggers). 
A schematic of the event trigger is shown in Figure
\ref{fig:function.ps}.

\subsubsection{Discriminator Efficiency}
As the first element in the event trigger, it is essential that the
Lecroy 3140 Discriminator module operates efficiently for nanosecond
duration pulses.
The discriminator efficiency is checked explicitly in the lab using
an 800 ps width signal (generated by an HP8133A signal generator) 
fed directly into a 3140 Discriminator
module, and also verified in the field by pulsing the array with
a 600 ps width signal broadcast through one of the broadband TEM horns
(the bandwidth of the horns is $\sim$3 GHz, so timing is preserved 
down to $\sim$300 ps). 
The LeCroy discriminator is found to be $>$99\% efficient in triggering
on the narrowest width pulses our signal generator is capable of producing.
Given the
measured impedances of our cable, amplifiers, and antennas (discussed
below),
a neutrino pulse is expected to produce signals of width $\sim$1-3 ns\cite{fmr}
at the surface.

\subsection{Raw Data}
Raw trigger rates (before veto) are typically 30 Hz;
data-taking rates after the veto are typically 0.01 -- 0.1 Hz.
The data-taking rate (and, correspondingly, our 
experimental livetime) is limited by:
a) the time required to write information from the digital 
oscilloscopes to
disk ($\sim$10 sec/event), b) the time required to 
perform the surface-background veto in software ($\sim$10 msec/event), and
c) our inability to take data at those times when the South Pole satellite
uplink (broadcast at 303 MHz) is active, due to the high amplitude of the
uplink signal.\footnote{Notch filters
will be installed in 2002 to mitigate this background.} 
When this uplink is not active, the discriminator
thresholds correspond to typical livetimes of $\sim$80\%.
Based on known dead times in the system, integrated array on-time is
currently monitored and updated on-line.

\begin{figure}[htpb]
\includegraphics[width=6in,angle=0]{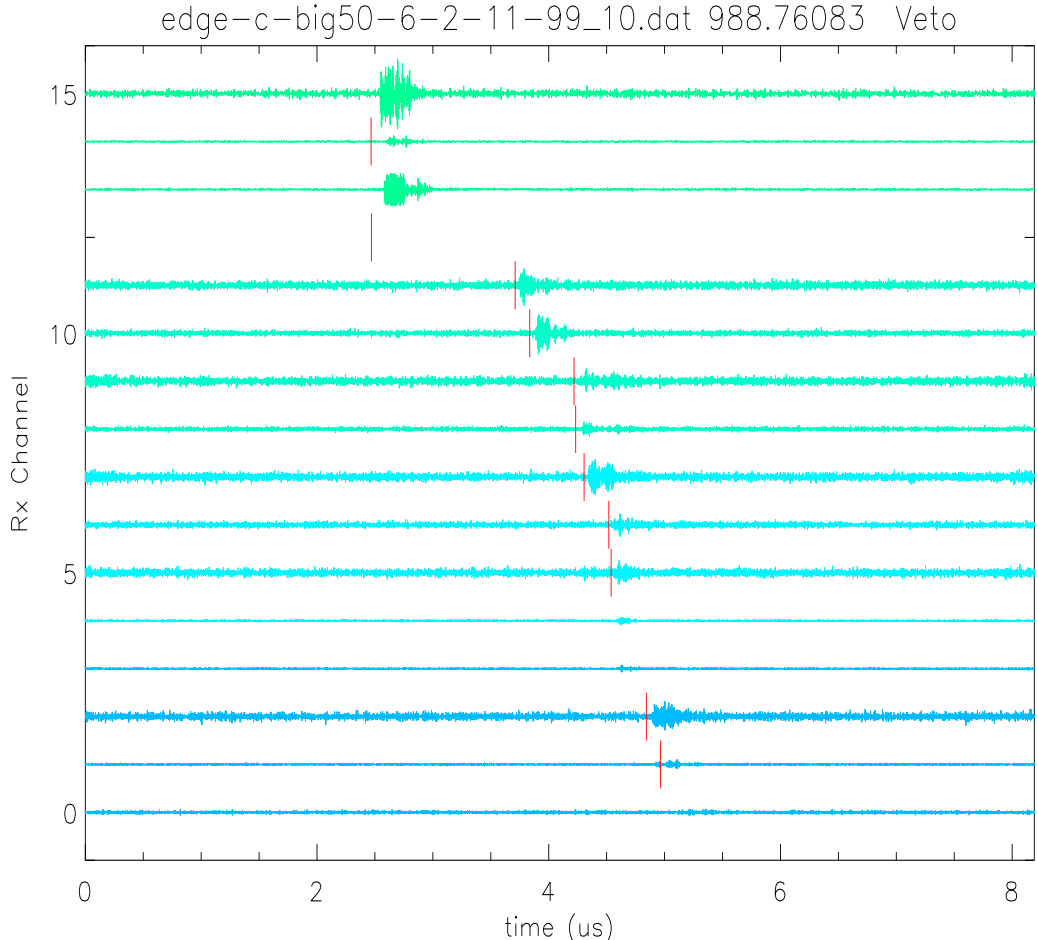}
\vspace{0.5cm}
\caption{Typical data event recorded by the RICE array (Feb. 11, 1999).
This event was vetoed as having a surface origin, based
on the observed sequence of hit times (see text).}
\label{fig:surface_veto}
\end{figure}

A typical event, as recorded in 
the HP54542A oscilloscopes, is shown in Figure
\ref{fig:surface_veto}. 
In the event display, each of the 8.912 $\mu$sec traces corresponds to one
of the under-ice antennas. 
Vertical bars indicate the TDC time recorded
for each particular channel. Note that
one channel (12) has no waveform information, and two channels
are not used in the trigger (13 and 15) and therefore
have no TDC information. Also
note that channels which do not have hits which exceed the
initial discriminator threshold do not register a TDC time.
The rise time and leading-edge
timing resolution of each
receiver as determined by waveform information can be 
approximately estimated from
the Figure. 
The leading edge resolution is typically 2 nsec; the ring time
for each antenna is typically tens of nanoseconds.
In the Figure, the antennas have been ordered according to their
distance from the surface, with the shallowest (/deepest)
antennas at the top (/bottom) of the Figure. The pattern of hits
shown is therefore easily identifiable as a surface-generated 
RF pulse sweeping down through the radio array.

\begin{figure}[htpb]
\includegraphics[width=4in,angle=90]{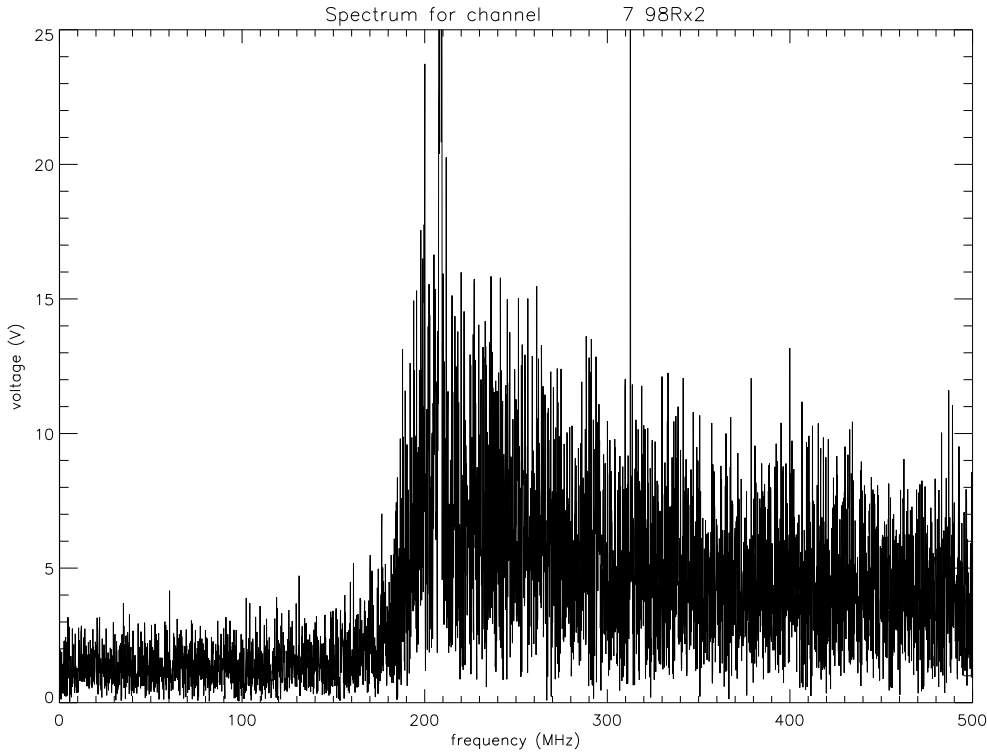}
\caption{Fourier transform of raw data waveform corresponding to
channel 7 of the previous Figure.
The South Pole station
satellite uplink at 303 MHz is evident in the Figure. 
A high pass filter suppresses signals below $\sim$180 MHz.
Other features of
this distribution are discussed in the text. No corrections for cable loss,
signal amplification, etc. have been made. 
(The vertical scale is arbitrary.)}
\label{fig:FFT}
\end{figure}

The Fourier transform of a
typical waveform signal (channel 7) is
displayed in Figure \ref{fig:FFT}. The general shape of the signal in
the frequency domain reveals many interesting features of the experimental
hardware and the environment. Below 200 MHz, suppression of noise due to the
filter is evident; as mentioned previously, 
use of this filter was motivated by the need to reduce
low-frequency RF backgrounds due to 
broadband galactic sky noise,
monochromatic
Continuous Wave (CW) sources at the Pole,
as well as the
firing of AMANDA phototubes.
Above 200
MHz, losses due to cable effects become increasingly pronounced.
At the time this event was recorded, the 303 MHz 
satellite uplink at the Pole was active;\footnote{This event
was taken before our ``303-veto'' was
functional.} the peak at this frequency is
evident in this Fourier Transform. 

\section{Timing Calibration}
\begin{figure}[htpb]
\centerline{\includegraphics[width=10cm]{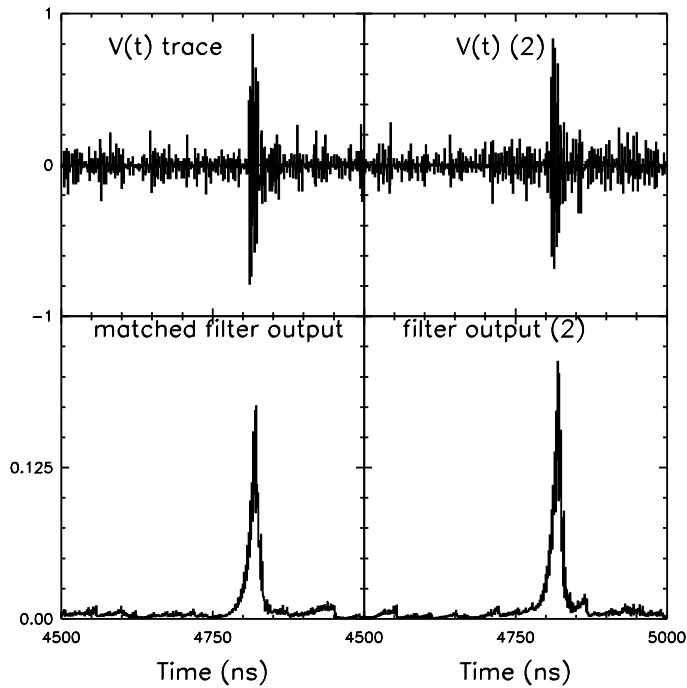}}
\caption {Raw V(t) for two waveforms (top); the signal time
as determined by a simple maximum voltage algorithm corresponds to the
maximum excursion from zero. Bottom displays show the quantity
$\Sigma f(t)V(t)$, with $f(t)$ being the damped oscillator
reference signal and $V(t)$ the actual recorded waveform shown
in top panels. The signal to noise is clearly greater
for the filter output.}
\label{fig:V-vs-filter.ps}
\end{figure}

Event and source reconstruction is based on our knowledge
of the array geometry, ice properties and thus the expected
times for a wave-front to propagate from the source to any given
receiver location. A time-of-hit is defined as the time of
the first excursion exceeding $6\sigma_{rms}$ in a waveform;
$\sigma_{rms}$ is determined from the sample of ``unbiased'' events
which are taken every 600 seconds, independent of the
event trigger status. The maximum resolution on 
the hit-times cannot, therefore, exceed the sampling time of the
oscilloscopes (1 ns). This resolution can be improved using a
``matched filter'' algorithm which matches the observed
waveform with a reference signal waveform; such a software 
algorithm is currently under development. In the ``matched filter''
approach, one defines a reference signal corresponding to the
expected damped oscillator response of the antenna to an impulsive 
signal: $f(t)\sim e^{-Rt/L}cos(\omega t)$, with $\omega=\sqrt{1/RC}$.
The values of $R$, $L$ and $C$ can be estimated from network analyzer
measurements of the complex 
antenna impedance ${\vec Z}_{antenna}$ (described below), 
as a function of frequency. The
signal time is defined as that time which maximizes the integral:
$\int f(t)V(t)dt$, with $V(t)$ given by the direct data measurement of
the waveform voltage with time.
As an illustration of 
the potential improvement arising from the matched filter algorithm,
Figure \ref{fig:V-vs-filter.ps} compares the raw signal $V(t)$ 
(top panels) with
the output of the filter algorithm (lower
panels) for one channel (channel 2) in one
event. The signal to noise is clearly
superior in the latter case.

Knowing the time differences $\delta t_{ij}$ between 
all pairs $(i,j)$ of hit antennas, we perform $\chi^2$ 
minimization to
find a source location. 
Once the vertex location has been found,
superimposing a Cherenkov cone of
half-width 57$^\circ$\footnote{$n_{ice}$(200 MHz)$\sim$1.78.}
on the hit antennas allows a 
determination of the source direction. For the full
reconstruction, this method requires at least four antennas to be
hit (i.e., 3$\delta t_{ij}$ values). 
Uncertainties in $\delta t_{ij}$
arise from several sources, including
risetime resolutions ($\sim$2 ns),
differences in signal propagation velocity in the ice due to variations
in the dielectric constant with depth, 
differences in signal propagation speed 
within the different analog cables being used, differences in cable lengths, 
and receiver deployment surveying uncertainties.

Buried transmitters 
(``Tx'') are used to calibrate the channel-to-channel timing
delays. A short duration pulse is sent to one of
the five under-ice transmitters, which
subsequently broadcasts the signal to the receiver array. 
An event vertex
is reconstructed exclusively from the measured 
channel-to-channel time delays; constraining the source to a
unique location
allows a calculation of the timing
residual $\chi^2$ for each channel, based on the timing uncertainty:
$\chi^2_t(ij)=({\delta t_{ij}^{\rm measured}-\delta t_{ij}^{\rm
expected}\over\sigma_t(ij)})^2$.
An iterative procedure is
used to calibrate out the
observed channel-to-channel timing delays and minimize the
timing residuals for an ensemble of events. 
Typical timing
calibration corrections are $\sim$10 ns per channel; these
corrections are then used for all subsequent event reconstruction.

\begin{figure}[thpb]\includegraphics[width=16cm]{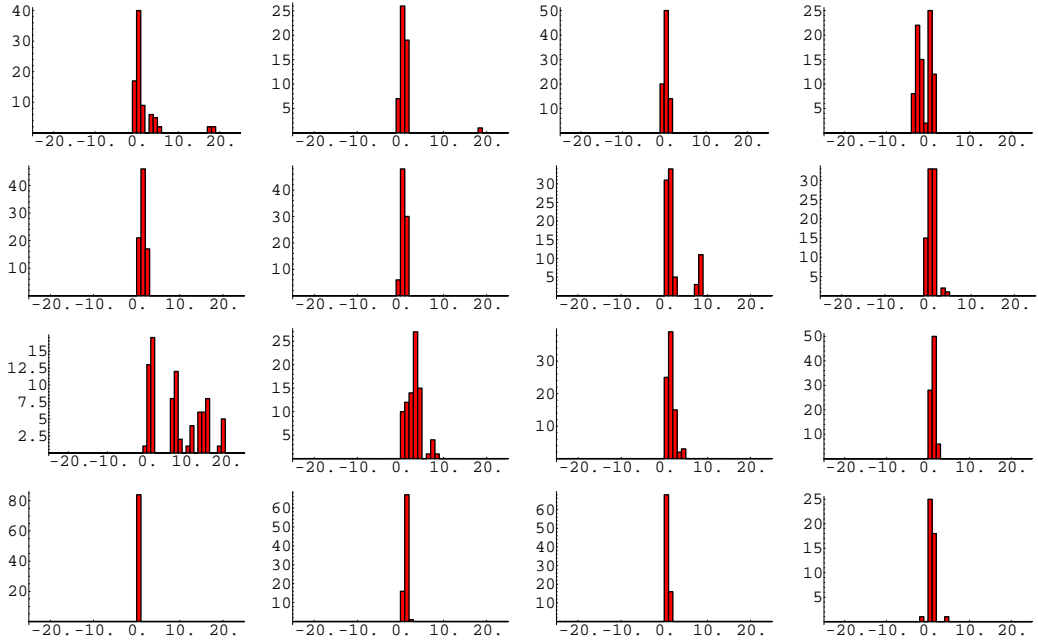}
\caption{Time difference between calibration signal arrival time in 
ch. i relative to ch. 12, in nanoseconds.
Data taken from transmitter pulser event sample. Oscilloscope jitter
in channel 8 is evident from the distributions.}
\label{fig:ptfig.eps}\end{figure}

Once the source location has been determined,
the timing resolution for each channel can be derived by
examining the width of the 
$(\delta t_{ij}^{\rm measured}-\delta t_{ij}^{\rm expected})$ 
distribution.
After time calibrations have been performed,
these differences of
time differences should (ideally) be zero for a unique vertex. 
Figure
\ref{fig:ptfig.eps} shows these distributions,
with channel 12 (arbitrarily) chosen as the
reference channel.
From these plots, the signal arrival
time resolution for good channels (i.e., all channels 
except channel 8 in the Figure)
is determined to be $\sim$1.5--2 ns, slightly
larger than the oscilloscope sampling time of 1 ns.



\subsection{Vertex Reconstruction Algorithms}
Two algorithms 
are used to determine a
source location. 
The first, 
a ``trial-and-error'' procedure, tests the consistency
of each possible (x,y,z) source location, in a grid 2km x 2km x 1km
below and around the radio array,
with the observed data. I.e., the $(xgrid,ygrid,zgrid)$ 
source point (using a step size 
of 10 m, or a total of $4\times 10^6$ source points tested) corresponding
to the minimum $\chi^2$ is identified. The advantage of this procedure 
is that time distortion effects due to
ray tracing through regions with varying
index of refraction can easily be incorporated; the disadvantage is 
speed and the intrinsic dependence on the grid spacing. 
The second procedure analytically
determines the source location 
${\vec r_{source}}$ and a global event time $t_0$ by simultaneous
solution of the 4 equations: 
$|r_{Rx,i}-r_{source}|=(c/n)t_i$ ($i=1,2,3,4$), where
$r_{Rx,i}$ is the vector from the 
origin to receiver $i$, $r_{source}$ is the vector from the origin to the
source, and $t_i$ is the 
hit time recorded for the $i^{th}$ receiver.
There are at most two roots: eliminating complex or acausal
roots, while also
requiring consistency of 5 or more hits resolves
ambiguities.
In Cartesian coordinates, we
denote this analytic solution for the source 
location as ${\vec r}_{source}$=$(x4hit,y4hit,z4hit)$.

\begin{figure}[th]
\centerline{\includegraphics[width=12cm]{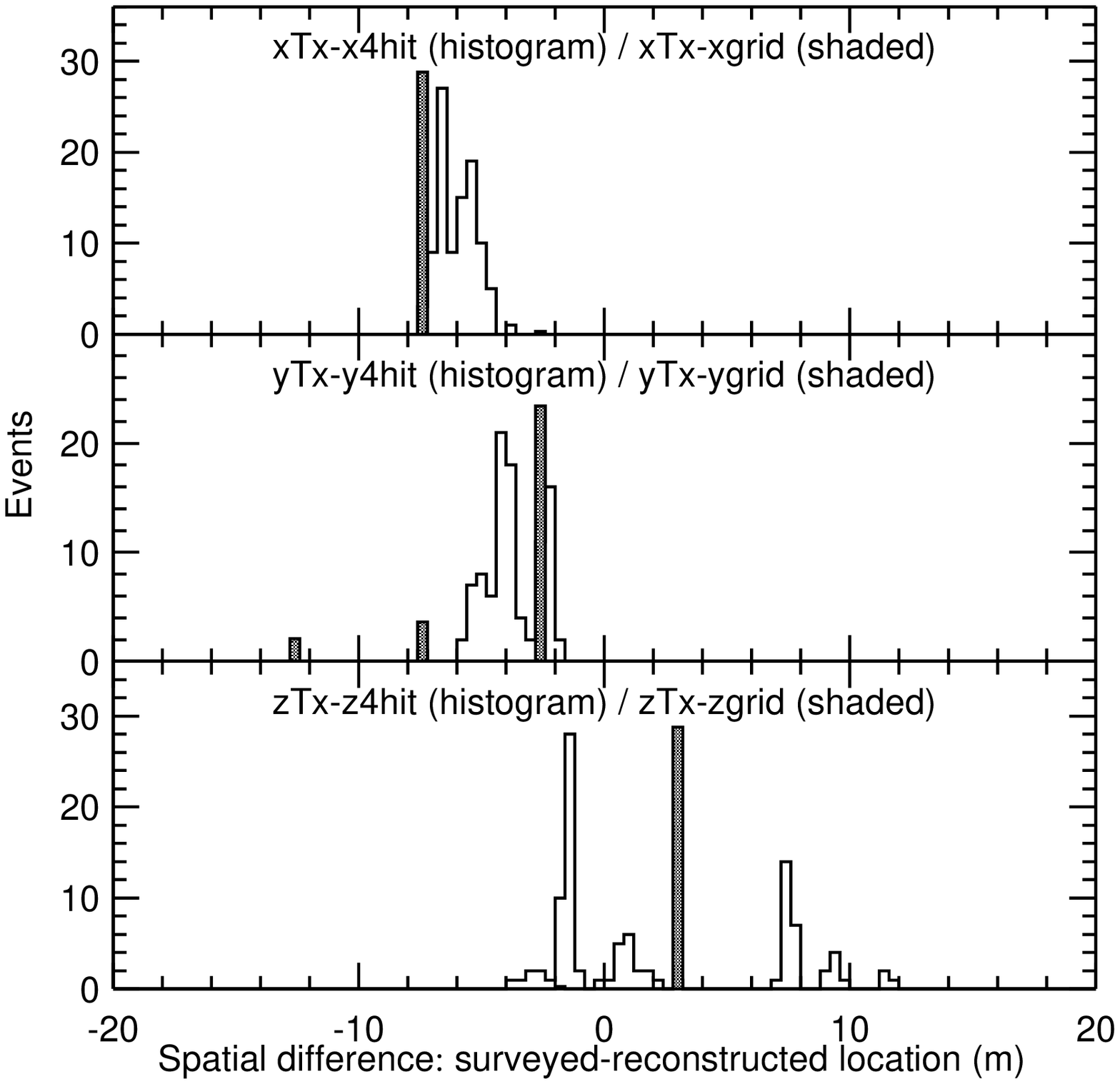}}
\caption{Reconstruction 
of Tx pulse from transmitter
97Tx2, based exclusively on measured receiver times. 4-hit solutions 
are shown as the unshaded histogram; grid solutions to the vertex location
are shown as the shaded histogram.}
\label{fig:Tx_recon}
\end{figure}
Source reconstruction results are shown in 
Figure \ref{fig:Tx_recon}. 
Three plots,
each showing the difference between the reconstructed
vertex location and the surveyed vertex location in one of the
spatial coordinates, are shown. Based on the (general) consistency
between both the 4-hit and the grid algorithms with
the surveyed Tx location, we conclude that the 
surveying error in the Tx location is less than 5 m.

\begin{figure}[htpb]
\centerline{\includegraphics[width=5in,angle=0]{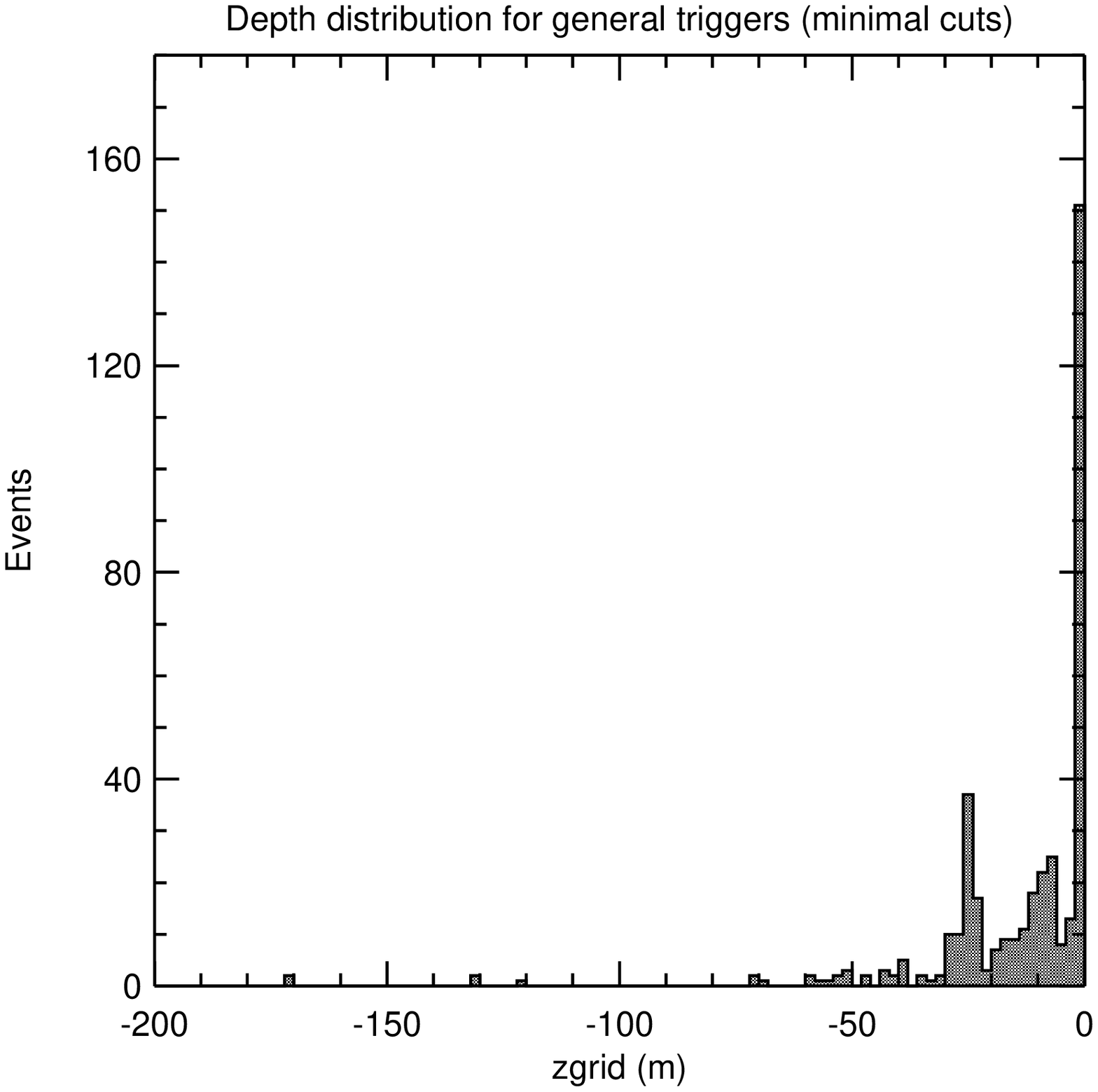}}
\vspace{0.5cm}
\caption{Typical depth distribution for 
a sample of 348 general RICE triggers. Triggers
are observed to be dominated by sources near the surface. (There are no
underflow entries.)}
\label{fig:gen_trigs_z0fit.ps}
\end{figure}

For general events (such as those displayed in Figure \ref{fig:surface_veto}),
we expect backgrounds generated at the surface to dominate.
Figure \ref{fig:gen_trigs_z0fit.ps} shows the reconstructed source depth
($zgrid$) for a sample of general triggers; as expected, sources 
populate the region above the array.\footnote{Note that the
$z=0$ bin includes all source
depths corresponding to $z>0$, as well.}
\vspace{1cm}

\subsection{Dielectric Constant of Ice}
\begin{figure}[htpb]
\centerline{\includegraphics[width=10cm]{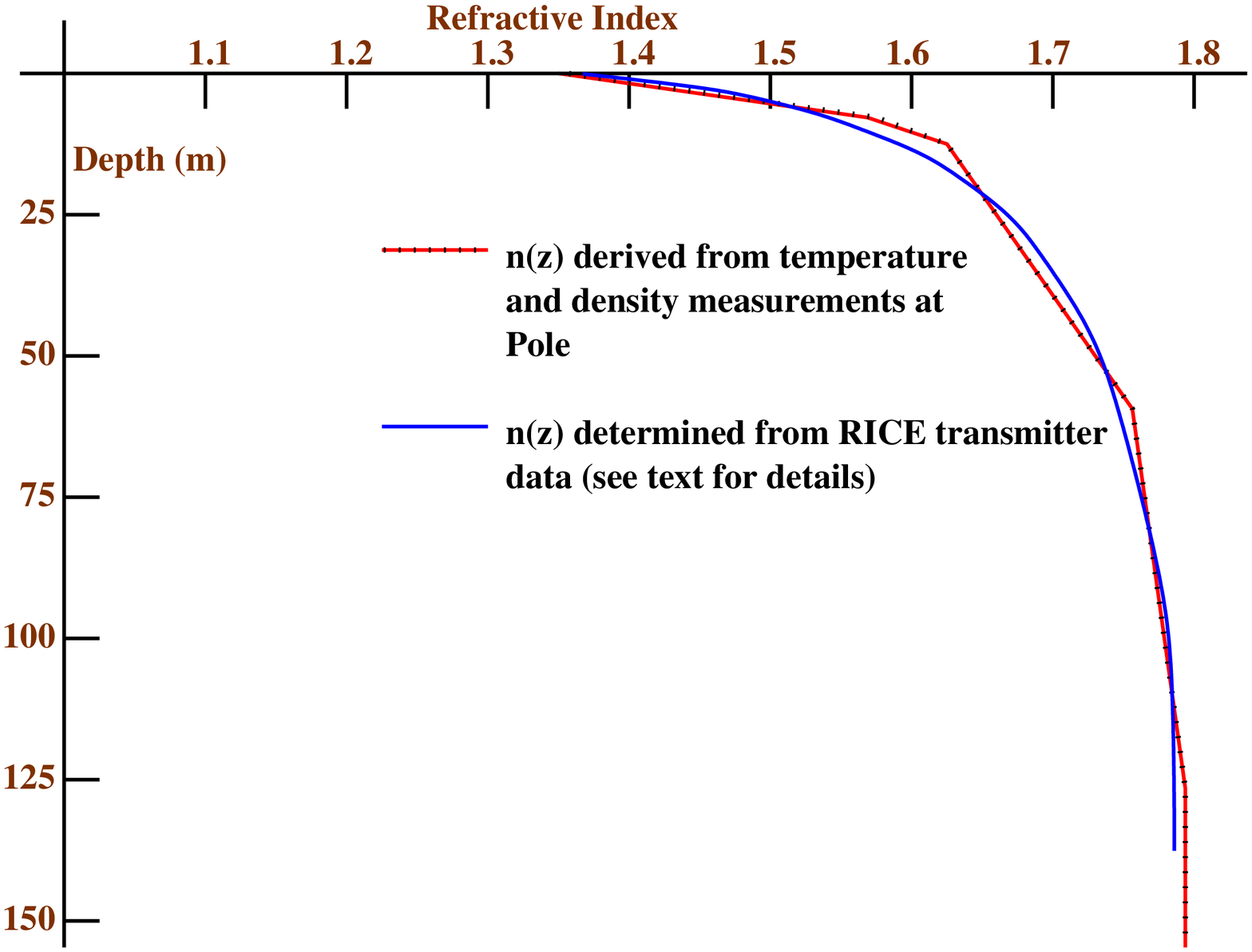}}
\caption{Index of refraction as a function of depth.}
\label{fig:n(z)}
\end{figure}

The complex dielectric constant ${\epsilon}(\omega)$ allows
one to calculate both the absorptive and refractive effects of the ice.
The imaginary part of the dielectric constant 
(related to the ``loss tangent'') 
prescribes losses due to absorption; the real part corresponds to
the refractive index of the medium. 
Information on the refractive index can be derived from 
temperature and density profiles,
as a function of depth,
acquired by AMANDA deep drilling operations.
Such profiles ($z(T,\rho)$) 
can be combined with laboratory measurements of the
dependence of the index of refraction of ice on temperature and 
density ($n(T,\rho)$) to predict the expected index of refraction
profile at the South Pole as a function of depth $n(z)$.
This derived $n(z)$
function can be compared with the profile calculated directly
from RICE transmitter data. Using the surveyed transmitter and receiver
locations, combined with Fermat's principle, one can determine the
profile $n(z)$ which best reproduces the observed $Tx\to Rx$ 
radio signal transit times.
A preliminary 
comparison between the measured $n(z)$ function determined from
transmitter data with the ``derived'' $n(z)$ profile is shown in
Figure \ref{fig:n(z)}. Agreement is satisfactory; 
however, work is still in
progress to quantify the agreement between the
two curves.

Absorption of radio waves in cold glacial ice is 
temperature, density, 
and frequency dependent and has been measured \cite{Bog80},
indicating attenuation lengths $\alpha\sim$1-10 km at
frequencies in the range 100 MHz -- 1 GHz.
Given the small scale of our current array
($\sim$100 m) compared to the very large attenuation length 
$\alpha$
expected for
Polar ice in the 100 MHz -- 1 GHz regime, our
transmitter tests are consistent with zero absorption over our short
baseline; 
long baseline
transmitter measurements are an objective of future Polar
campaigns.

\section{Antenna Response and Gain Calibration}
To detect a neutrino of a given energy
interacting at a given distance from an antenna,
we need to quantify the sensitivity of the array to the resulting
Cherenkov signal.
Although neutrino vertex location reconstruction is based only on 
channel-to-channel timing,
an amplitude calibration is needed
in order to ensure that the discriminator efficiency, for
a given incident neutrino energy,
is reliably calculable.

\subsection{Antenna Effective Area}
One of the fundamental parameters used to define
the power response of an antenna is the effective 
area $A_{eff}(\theta,\phi)$ (or gain $G=4\pi A_{eff}/\lambda^2$); 
this is related to the incident signal intensity $I$ as:
$P_{out}=IA_{eff}$. 
We have used two techniques to measure the effective
area $A_{eff}$ of RICE dipole antennas. The first applies the
Friis Transformation
Equation to data taken when a transmitter
(Tx) broadcasts to a receiver (Rx)
on the KU Antenna Testing Range (KUATR).
The Friis Equation relates the power
broadcast by the transmitter to the signal intensity measured
at the Rx (valid in the far-field case). Defining the
monochromatic power
into the transmitter $P_{Tx,in}$, the power broadcast outwards from the
transmitter $P_{Tx,out}$, the
power intercepted by the receiver $P_{Rx,in}$,
and the power transmitted out the back of the receiver: $P_{Rx,out}$ (which
we measure either on an oscilloscope or a network analyzer),
we have:
$P_{Tx,out}=P_{Tx,in}G_{Tx}$, $I_{Rx,in}=P_{Tx,out}/(4\pi R^2)$,
$P_{Rx,out}=I_{Rx,in}A_{eff,Rx}=P_{Tx,out}A_{eff,Rx}/(4\pi R^2)$;
${P_{Rx,out}\over P_{Tx,in}}=G_{Tx}A_{eff,Rx}/(4\pi R^2)$.
Using a calibrated transmitter, $G_{Tx}$ is known; by comparing to 
a calibrated receiver ($A_{eff,Rx}^{standard}$ known), the gain or
effective area of any arbitrary antenna can be inferred by a simple
ratio. 

Alternately, the gain/effective area can be determined by
sending a known amount of power through a cable and into an
antenna. The reflection coefficient $\Gamma_R$ (having real and imaginary
components $\Gamma_r$ and $\Gamma_i$, respectively) measures the
impedance mismatch of the antenna ${vec Z}_{antenna}$
and the cable $Z_{cable}$, and is ideally zero for a high-gain
antenna (impedance well-matched to 50$\Omega$ cable, e.g.). Numerically,
$G\sim (1-|\Gamma_R|^2)$. The two measurements of $A_{eff}$ 
(from the testing
range and the impedance measurement, respectively) agree to within $\pm$1 dB
over the frequency range of interest and are consistent with
simple expectations for dipoles.

\subsection{Effective Height}
To understand the shape of the signal voltage produced by a
receiver in the time domain, we must quantify the complex effective height
${\vec h}$\cite{Har65}.
The effective height ${\vec h}$ (in units of meters)
is related to 
the magnitude and the phase of the voltage
resulting
from the application of a complex 
electric field vector 
at the antenna load by:
$V_{out}={\vec E_{in}}\cdot{\vec h}$ =
$h {\vec E_{in}} \cdot {\hat n_A}$.
The effective area and the magnitude of the 
effective height can
be related through: 
$|{\vec h}|\sim\sqrt{A_{eff}/(120\pi)}$ 
$=\sqrt{\lambda^2 (Gain)/(480\pi^2)}$.
The polarization of ${\vec h}$ is aligned along the 
dipole axis 
${\hat n}_A$ as ${\vec h}=h{\hat n}_A$, 
where $|h|$ is the magnitude of effective height.

The full,
complex transfer function ${\vec T}(\omega)$
for the antenna, in principle, gives a complete
description of the antenna (+cable) 
response and can be related to the
effective height. This function ${\vec T}(\omega)$ can
be written as the product of the complex impedance of the 
(antenna+cable)
${\vec Z}(\omega)$ multiplied by the complex height function 
${\vec h}(\omega)$, properly taking into account potential
mismatches between the impedance of the antenna and
the impedance of the attached cable:
${\vec T}={\vec h}(Z_{cable})/({\vec Z_{antenna}} + Z_{cable})$,
with
$Z_{cable}$=50$\Omega$+$i0\Omega$.
Both the magnitude and
phase of the effective height are determined directly from KUATR
measurements.
${\vec Z_{antenna}}$ is determined from reflected power
measurements on a HP8713C Network Analyzer (NWA).
An independent check on the internal consistency of our
${\vec Z}_{antenna}$ determinations is presented in Appendix 1.

\begin{figure}[htpb]
\begin{picture}(250,300)\includegraphics{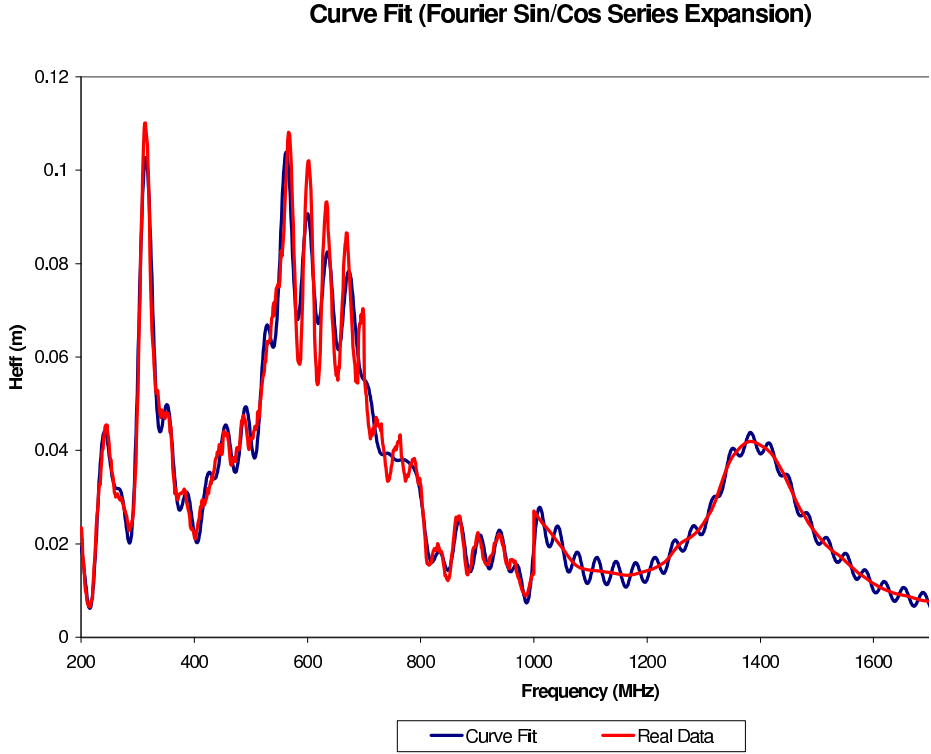} \end{picture} 
\caption{Modulus of RICE dipole effective height (m.) as a 
function of frequency. Shown are data taken on the KU antenna
testing range, with a Fourier fit overlaid.}
\label{fig:heff.ps}
\end{figure}

The magnitude of the effective height measured for a typical
RICE dipole
is given in Figure \ref{fig:heff.ps}. The peak frequency
in air\footnote{This peak frequency is
shifted down in ice by $n(\omega)$=1.78 at radio-frequencies.}
($\sim$600 MHz)
is roughly consistent with the dimensions of the half-dipole ($\sim$15 cm);
the effective height, as expected, also has magnitude of order 10 cm. at 
the peak frequency.

\begin{figure}[htpb]
\centerline{\includegraphics[width=11cm,
angle=90]{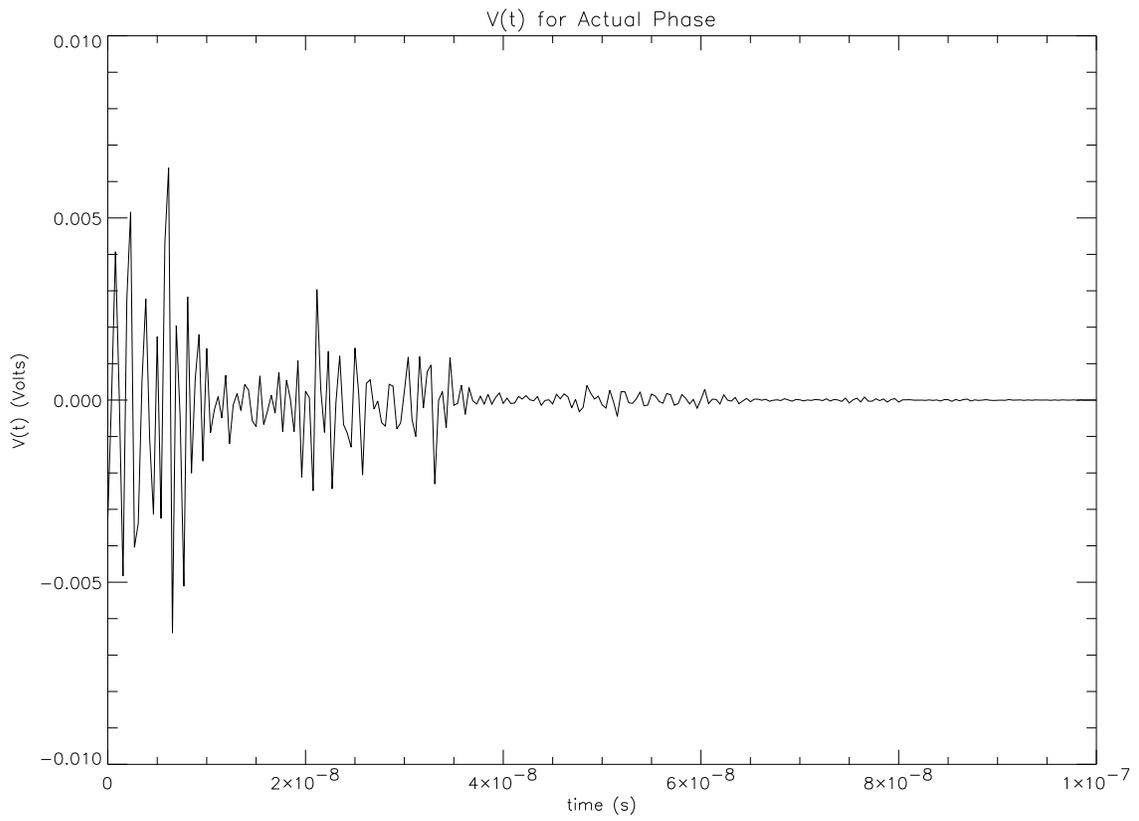}}
\caption{Expectation for observed
RICE receiver signal shape $V'(t)$ based on measured
effective height and complex impedance function, given a short duration
input signal to a RICE transmitter.}
\label{fig:vtrealphase}
\end{figure}

The phase variation of the effective height has also been measured
as a function of frequency at KUATR and found to be negligible
(put another way, we find that, to a very good approximation the 
experimental slope of the phase variation with frequency is linear:
$\Delta\phi(\phi)\approx 0.031\phi$ (rad). Such a linear
variation has no net effect on
overall antenna response.).
Given the magnitude of the effective height, the phase variation of the
effective height, and therefore the magnitude and phase variation of the
complex antenna impedance ${\vec Z}(\omega)$, the
complex transfer function ${\vec T}(\omega)$ can now be calculated,
and used to
predict the expected waveform $V'(t)$ observed in a RICE antenna in response to
a transmitter signal. This is done by transforming the input impulse
$V(t)$ to frequency space, multiplying the function $V(\omega)$ by the
transfer function
${\vec T}(\omega)$ for both Tx and
Rx (properly normalized), and then transforming back to
the time domain to give $V'(t)$.
Figure \ref{fig:vtrealphase} shows the result of this exercise.
Qualitatively, the after-pulsing observed in data
is reproduced by our complex transfer function.
We note that the actual expected signal
shape observed for a neutrino event
requires knowledge of the details of the input 
signal, as discussed in the
references\cite{ZHS,BigPaper}.

Dipole response has 
also been measured as a function of both azimuthal and polar
angles. The polar angle response is observed to be well approximated by
a $cos^2\theta$ dependence (in power); the dipole response in azimuth is
observed to be flat, as expected.

\subsection{Amplifier Gain Calibration}
Signals from the antennas are boosted by two stages of amplification, 
totaling
between +88 dB and +96 dB of gain, depending on the channel. 
For a RICE waveform containing only
(``unbiased'') thermal noise,
the total power in a 
frequency bandwidth $B$ can
be calculated from the discrete
Fourier transform of the waveform as
$P_{noise}=kTB$ (we check several 50 MHz-wide frequency bins 
from
250 to 500 MHz for this calculation).
Since the total noise power in this band
at the input to the antenna can 
also be written as a sum over the rms voltage measured in each
frequency bin: 
$P_{<V>}=\Sigma_\omega {V_\omega^2\over Z}$,
we can 
also write $P_{<V>}=P_{noise}G$, where $G$ is the overall gain of the
system. Thus,
based on the rms voltage $<V>$
of the 8192 samples contained in these 
``unbiased'' waveforms, the gain of the amplifiers
can be calculated {\it in situ}. 
The amplifier gain measured this way is flat up to 
the bandwidth limit of the oscilloscopes (500 MHz);
direct laboratory measurements of the amplifier gain using
a network analyzer are consistent with flat 
response up to 750 MHz.
Most($>$90\%) 
of the amplifiers are stable to 1-2 dB over the 
course of data-taking thus far analyzed.

\subsection{Full Circuit Amplitude Calibration}
\begin{figure*}[t]
\centerline{\includegraphics[width=10cm,angle=90]{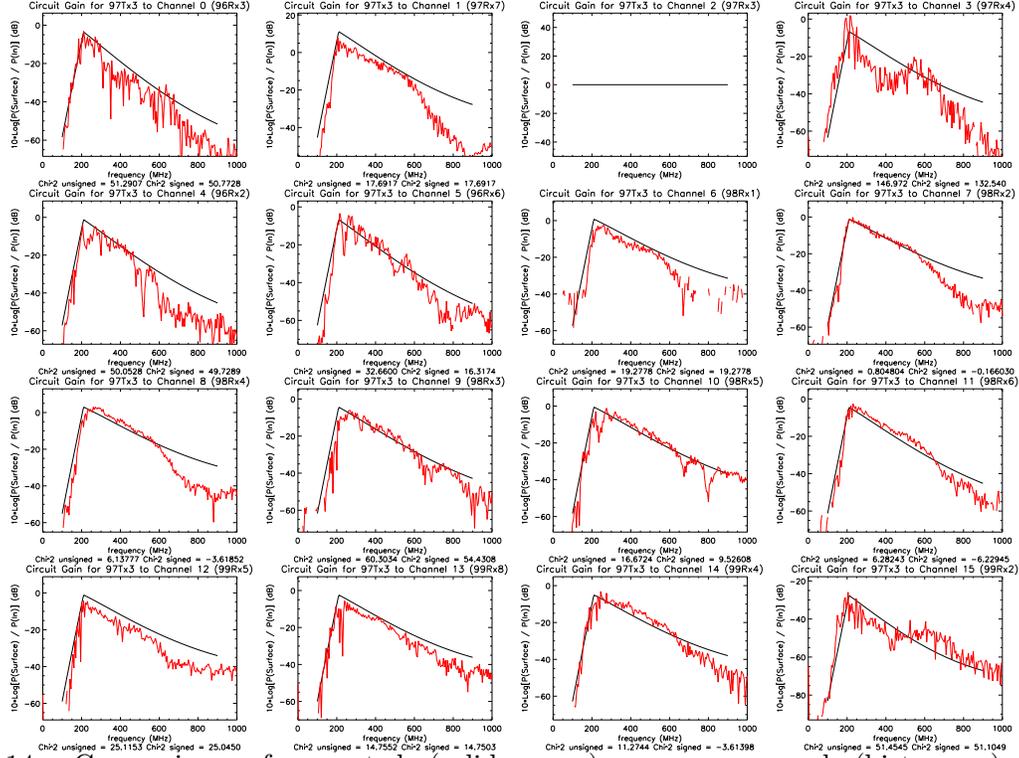}}
\caption{Comparison of expected (solid curve) vs. measured (histogram)
Tx$\to$Rx signal strength for one transmitter broadcasting to
16 receivers. Vertical scale is Return-Power/Transmit-Power, in dB.
No data is shown for the receiver (channel 2, top row) in the same hole
as the transmitter being used for this test, due to 
possible cross-talk effects. Corrections
for the measured roll-off of surface amplifier gain for some channels
above 700 MHz have been only approximated.}
\label{fig:circuit_gain}
\end{figure*}
In the final step of our amplitude calibration, the
antenna response to a continuous wave (CW) signal 
broadcast from an under-ice transmitter is measured {\it in situ}.
This test calibrates the combined effects of all cables, signal splitters,
amplifiers, etc. in the array.
A 1 milliwatt (0 dBm) 
continuous wave signal is broadcast through the transmit port of
an HP8713C NWA. The NWA scans through the frequency range
0$\to$1000 MHz in 1000 bins, producing a 0 dBm CW 
signal in each 
frequency bin. The signal 
is transmitted down through $\sim$1000 feet of coaxial cable to 
one of the five
under-ice dipole transmitting antennas. 
The transmitters subsequently broadcast this signal
to the under-ice receiver array, and the
return signal power from each of the receivers
(after amplification, passing upwards
through receiver cable and 
fed back into the return port of the NWA)
is then measured. Using laboratory measurements
made at KUATR of: 
a) the effective height of the dipole antennas, as a function of
frequency (previously described), b) the dipole Tx/Rx
efficiency as a function of polar angle and
azimuth, c) cable losses and dispersive effects
(cables are observed
to be non-dispersive for the lengths of cable, and over the frequency
range used in this experiment),
d) the gain of the two stages of amplification as determined from RICE
data acquired {\it in situ} by
normalizing to thermal noise
$P_{noise}=kTB=\Sigma_{\omega}<V_{ant}^2>/Z$, summing
over all frequency bins, and e) finally 
correcting for $1/r^2$ spherical spreading
of the signal power, one can model
the receiver array 
and calculate the expected signal strength returning
to the input port of the network analyzer.
This can then be directly compared with actual measurement.
Such a comparison, as a function of
frequency, is shown in Figure 
\ref{fig:circuit_gain} for one transmitter (97Tx3). 
Below 200 MHz, the attenuating effect of the high-pass filter is evident.
From the Figure, agreement is observed to be $\sim$3-6 dB (in power) for
the full circuit gain. Note that no correction for ice absorption has
been made, given the small scale of the array.\footnote{Nor have 
corrections been made for possible AMANDA cable ``shadowing'' in
the same ice-hole, which is
evidently not a significant effect.}

\begin{figure*}[thpb]
\centerline{\includegraphics[width=11cm,angle=0]{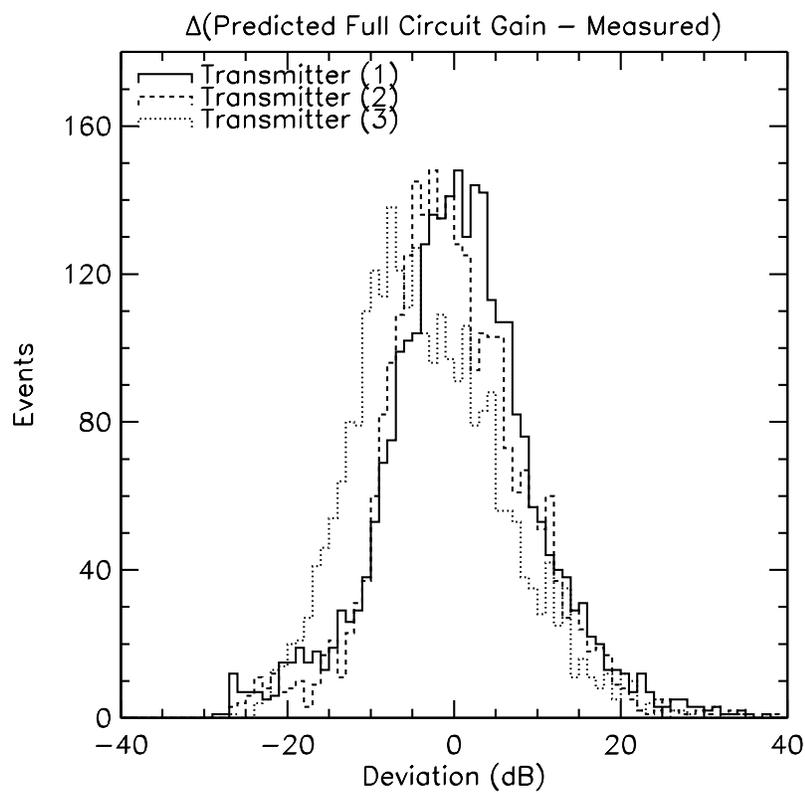}}
\caption{Deviation between expected vs. measured 
Tx$\to$Rx signal power for three transmitters broadcasting to
16 receivers.}
\label{fig:full_circuit_gain_dzb}
\end{figure*}

Similarly, for each frequency bin we can calculate the difference
between calculated vs. measured full-circuit gain.
Figure \ref{fig:full_circuit_gain_dzb} shows the deviation between
the calculated gain minus the measured gain, for several data runs.
Included in the Figure
are each of the 500 one MHz bins between 200 MHz and 700 
MHz, for three transmitters. 
This Figure therefore shows the average deviation between model 
and measurement over that frequency range; 
for the five currently functional
transmitters, the mean differences between the expected and
the measured gain\footnote{This is calculated as
$\Delta(G_{calc}-G_{meas})\pm\sigma_{G_{calc}-G_{meas}}$, where
$\Delta(G_{calc}-G_{meas})$ is the mean of each of the
distributions shown in the Figure, and
$\sigma_{G_{calc}-G_{meas}}$ is the error in the mean, given
by the r.m.s. of the distribution itself divided by the
number of points in each distribution.}
are $-0.6\pm0.6$, $-0.8\pm0.6$, $-2.3\pm0.5$, $-3.4\pm0.6$ and
$-2.8\pm0.6$ dB. 
Within the ``analysis''
frequency band of our experiment (200 MHz - 500 MHz), our
quoted level of
uncertainty in the total receiver 
circuit power is $\pm$6 dB; this
value is commensurate with the width of
the gain deviation distributions. On average, however,
the calculated gains quoted above are
well within these limits. Note also that, for all of the 
transmitters, the measured gain is actually higher than
expected. To calculate eventual upper limits on the
neutrino flux, however, we will use the more conservative calculated
full-circuit gain.

\section{Monte Carlo simulation of detector performance}
In order to check our understanding of the 
timing and gain calibrations of our experiment, we have written
a Monte Carlo simulation of the RICE array. This simulation allows
us to study the expected response of the radio receiver array to
either a Cherenkov signal (as generated by a true
$\nu_e N\to eN'$ charged-current event) or a random
noise coincidence such as those
expected to dominate our backgrounds. 
The simulation also checks our expected timing resolution 
(2 ns, as determined by measuring the channel-to-channel residuals for
pulsed transmitter events in data) as well as study the effects of
the overall gain uncertainty discussed previously (6 dB per channel, max,
obtained by examining the residuals between calculated vs. measured 
gain for continuous wave signals
from 200-700 MHz). 
Expected vertex (spatial) and angular (pointing) resolution, as
well as neutrino energy resolution can, in principle,
be assessed with the simulation.

\subsection{Monte Carlo Event Generation}
Neutrino interactions are generated uniformly in incident angle,
with vertices uniformly populating the region:
$|x|<$1~km, $|y|<$1~km and $0<z<$1~km.
After specifying a source location and incident
neutrino energy,
the simulation generates
receiver times (smeared by our timing uncertainty
of $\sim$2 ns) and voltages at the input to the antenna,
based on GEANT simulations of radio signals in ice\cite{BigPaper}.
In addition to the voltage due to the neutrino interaction, thermal
noise is also simulated at the input to the antenna.
Antenna response, as a function
of frequency, is modeled as described above; knowing the
amplifier gain and cable losses, the voltage
recorded at the surface is calculated, and smeared by the
gain uncertainty of $\pm$6 dB ($\pm$3 dB in voltage).
With simulated times and simulated voltages,
simulated events are then 
reconstructed with the same software used for data.

\subsection{Checks of the Simulation - MC vs. Data Tx Depth Reconstruction}
\begin{figure*}[htpb]
\centerline{\includegraphics*[width=11cm,angle=0]{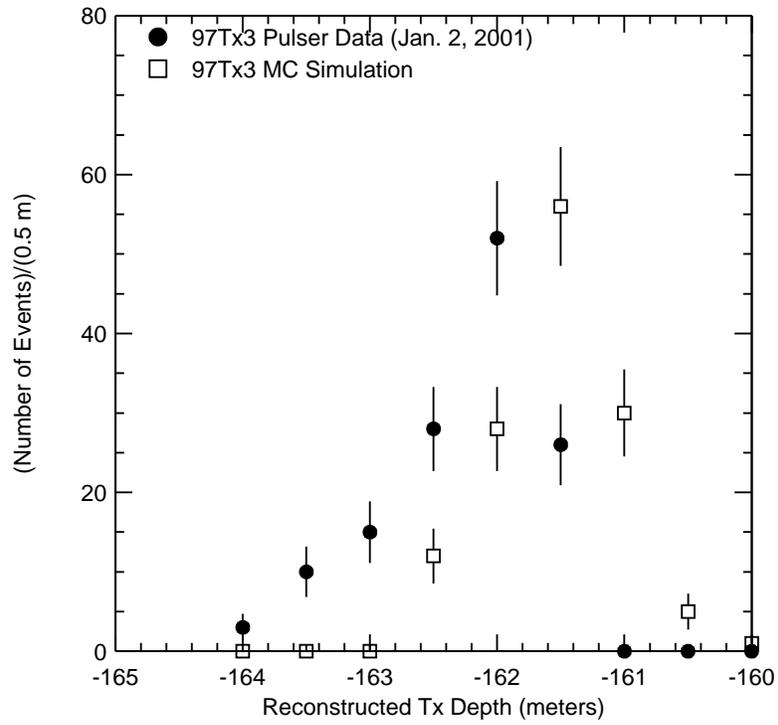}}
\caption{Comparison between reconstructed depth of transmitter (97Tx3), from
pulser data taken at the South Pole, and Monte Carlo simulations of
97Tx3 pulser events. Times
are smeared in the simulation by a Gaussian of width 2 ns.}
\label{fig:97Tx3_vtx_recon_MCvData.ps}
\end{figure*}

As a first check of the simulation,
we have compared reconstructed transmitter source depths in data vs.
Monte Carlo simulations. The source locations
reconstructed from data collected when
a pulser was connected to transmitter 97Tx3 were compared to 97Tx3
simulations, as shown in Figure \ref{fig:97Tx3_vtx_recon_MCvData.ps}.
Vertices are reconstructed using the analytic 
4-hit technique described previously.
The simulation adequately reproduces both the location (to
within 1 m) as well as
the width of the source distribution.

\subsection{Expected Vertex Depth Resolution Dependence on Depth}
\begin{figure*}[htpb]
\centerline{\includegraphics[width=11cm]{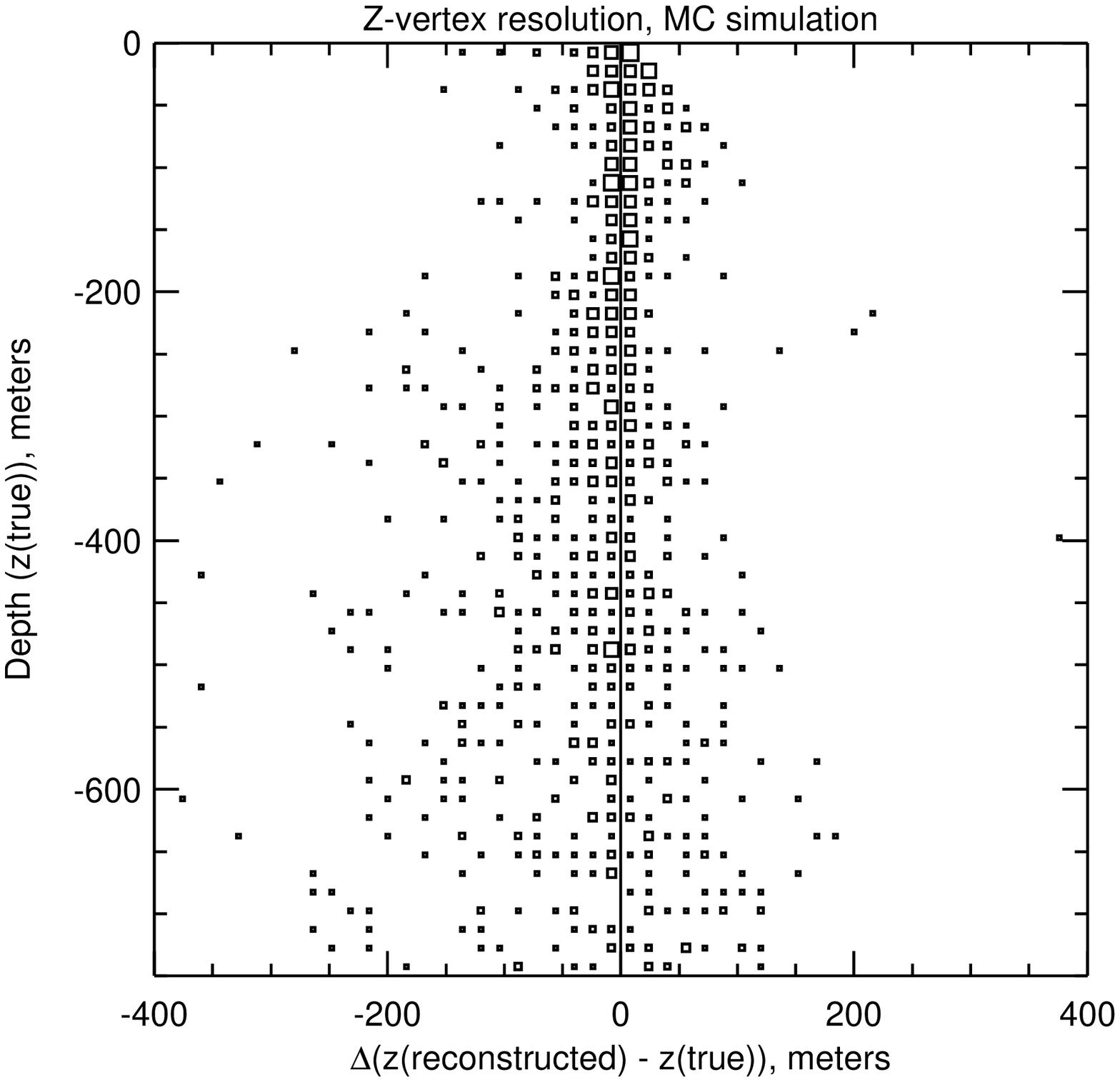}}
\caption{Deviation between reconstructed and true interaction depth
($z4hit$), vs.
true interaction depth, from Monte Carlo simulations. The reconstructed
vertex here was obtained using the 
analytic, 4-hit vertexing algorithm. The
size of the squares corresponds to the number of reconstructed events
in the Monte Carlo simulation.}
\label{fig:deltazVz.ps}
\end{figure*}

As a further example of the utility of the
simulation, Figure \ref{fig:deltazVz.ps} 
shows the expected vertex depth resolution (essential in discriminating
surface sources from in-ice sources), as a function of the true source
depth.
Not unexpectedly, the resolution is best when the 
source vertex is close to the array and the event geometry is best
resolved. Deep sources are increasingly difficult to pinpoint; additionally, 
the reconstruction software tends to reconstruct source vertices that are
closer than the actual source ($z_{reconstructed} - z_{true}<$0); i.e.,
vertices tend to be pulled closer to the array. This bias is under study.

\subsection{Angular Resolution}
\begin{figure*}[htpb]
\centerline{\includegraphics[width=12cm,angle=0]{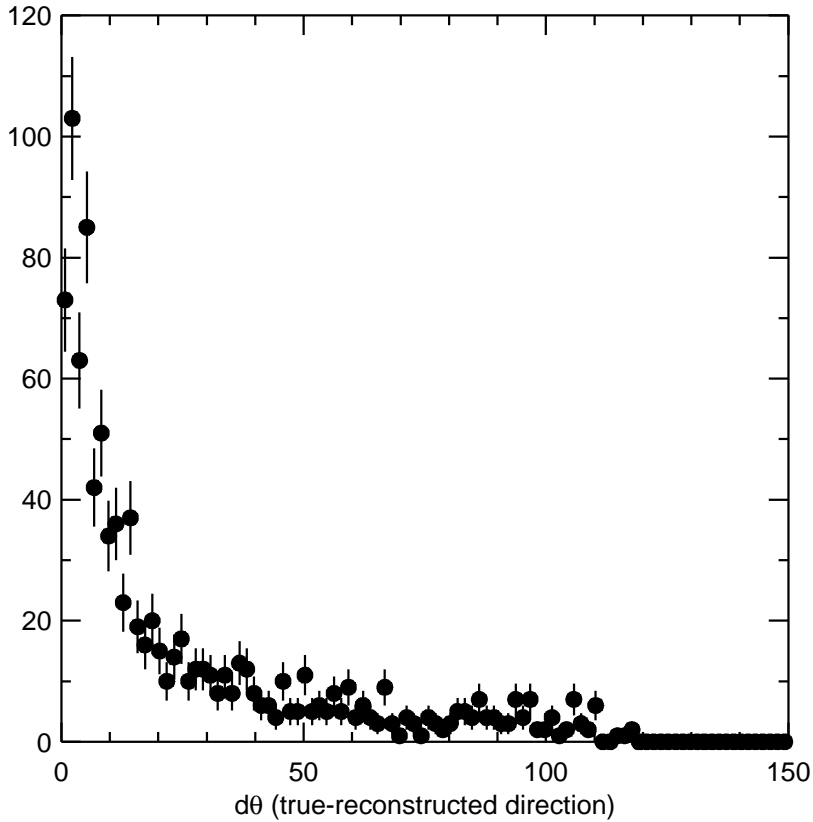}}
\caption{Monte Carlo prediction for
RICE array angular resolution, in units of degrees. 
Events ($E_\nu=$10 PeV) have been simulated over the
region within 1 km below and around the current array.}
\label{fig:angres.ps}
\end{figure*}

Good angular resolution is essential in discriminating
up-coming from down-going sources. Additionally, some
physics analyses (e.g., coincidences with Gamma-Ray Bursts) are limited by
the ability to point back to the recorded sky location of the GRB.
For neutrinos that interact very far from the array, the angular resolution
is expected to be poor -- the further the source point, the greater
the inability to measure the Cherenkov wavefront, and
the more the array ``looks'' like a single space-point relative to the
interaction point.
Figure \ref{fig:angres.ps} displays the angular resolution for
an ensemble of 10 
PeV neutrinos interacting within 1 km in the ice below the array;
we have required that at least four of
the simulated receiver voltages exceed threshold and can therefore be used in
vertex reconstruction.
For approximately half the events, the angular resolution is about 10 
degrees. The long tail in this distribution is due to distant source
points which have correspondingly
poorer pointing resolution.

\subsection{Energy Resolution}
\begin{figure*}[htpb]
\centerline{\includegraphics*[width=10cm]{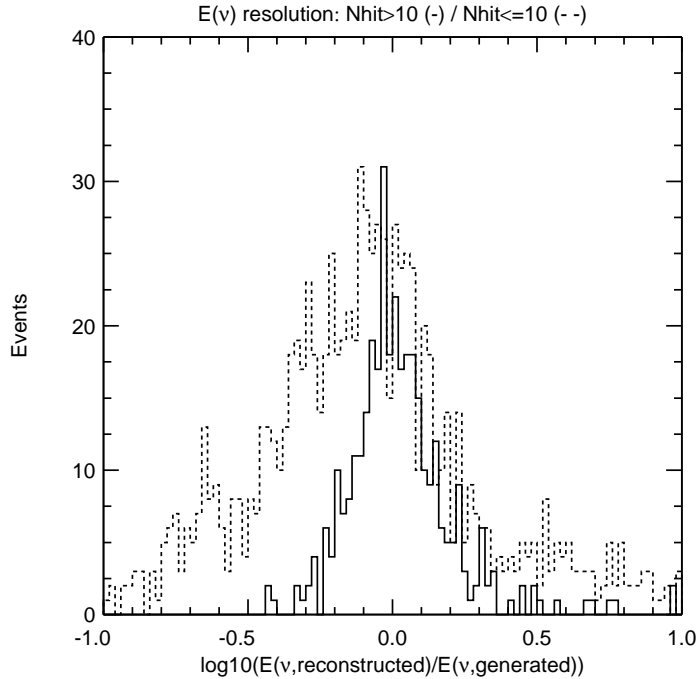}}
\caption{Histogram of
number of events vs.
logarithm (base 10) of Reconstructed/true neutrino energy, as
obtained from neutrino interaction simulations. 
The sample is divided into high hit multiplicities ($>$10) and low hit
multiplicities ($\le$10).
The energy resolution is
obviously better for events with higher hit multiplicities.
The simulated event sample is the same as that of the previous
Figure. }
\label{fig:Eres.eps}
\end{figure*}

Having reconstructed a vertex location, the distance from each antenna to
that vertex is determined. Having reconstructed the event geometry (i.e.,
the Cherenkov cone fit), we know the angular deviation
of each antenna off the cone. The greater this angular deviation,
the weaker the antenna signal strength.
Finally, given the voltages recorded on each
antenna, we have enough information
to make an estimate of the incident neutrino energy.
For each antenna, the inferred value of $E_\nu$ is calculated; we
define $E_{\nu,reconstructed}$ as the simple average of the
$E_\nu$ estimates obtained from each antenna.
Fig. \ref{fig:Eres.eps} shows the expected energy resolution
($log_{10}(E_{\nu,reconstructed}/E_{\nu,generated})$) 
for 10 PeV neutrinos generated
uniformly over distances within 1 km of the array. The resolution is obviously
superior in cases where the hit antenna multiplicity is large
($N_{hit}>$10) vs. cases where the hit multiplicity is small.\footnote{In 
making this plot, the voltages on the antennas were smeared by the
resolution on each channel (described previously); thermal
noise has also been simulated.}

\subsection{Conical vs. Spherical Source Event Geometries}
\begin{figure*}[htpb]
\centerline{\includegraphics[width=10cm]{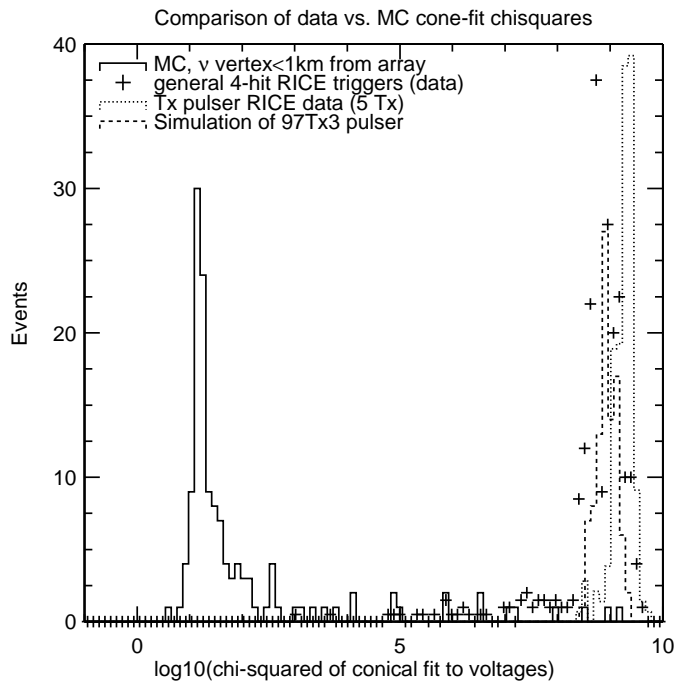}}
\caption{$log_{10}(\chi^2)$ of fit of recorded voltages (for both data
and simulation) for spherical vs. conical geometry sources. The separation
between conical and spherical sources, as well as the agreement between
simulation and data for spherical sources, is evident from the Figure.}
\label{alog10-MC-data.eps}
\end{figure*}

Perhaps most important in discriminating background from true
neutrino-induced events is the characteristic conical 
Cherenkov distribution of
energy in the latter case. Background (from the surface, e.g.) typically
consists of spherical waves due to transient sources. We measure the
consistency of a given event with either a conical or a spherical source
by a ``trial-and-error'' procedure similar to the ``grid''-based
vertex-finding algorithm: given a reconstructed source location, we find
the Cherenkov cone orientation most consistent
(in terms of a minimum $\chi^2_{cone}$ variable) with the observed
channel-to-channel voltages. Spherical sources should therefore
correspond to large values of $\chi^2_{cone}$ since a conical source
would typically have a smaller number of large-voltage signals than
a spherical source.
This is illustrated in Figure \ref{alog10-MC-data.eps}, where we compare
the $\chi^2_{cone}$ distribution for: a) general, 4-hit RICE triggers, taken
from data, b) calibration pulser events, taken from data, c) Monte Carlo
simulations of a pulser source (97Tx3) emitting spherical waves, and d)
Monte Carlo simulations of neutrino events producing Cherenkov cones.
The separation between conical vs. spherical source geometries is evident
from the Figure, as well as the rough agreement between data and the
simulation of a spherical source near the array.

\section{Summary}
Basic calibration
of both the time and amplitude response of RICE radio receivers has been
made, relying primarily on data taken {\it in situ}. 
A Monte Carlo simulation
has been written which reproduces the gross features of those
calibration data.
The calibration 
of the detector is sufficient to allow limits to be placed on the
incident high-energy neutrino flux.

\vspace{2cm}
\begin{acknowledgements}
We gratefully acknowledge the generous
logistical support of the AMANDA 
Collaboration (without whom this work would
not have been possible), 
the National Science Foundation Office of Polar Programs,
the University of Kansas,
the NSF EPSCoR Program, the University of Canterbury Marsden Foundation, 
and the Cottrell 
Research Corporation. 
Matt Peters of the U. of Texas SOAR group provided
essential
consultation on antenna design, calibration and
electrical engineering. 
The National Science Foundation's Research Experience for Undergraduates
Program provided support for
Jeff Allen (Shawnee Mission South High
School, currently at U. of California, Berkeley),
Eben Copple (KU. Physics Dept.),
Karl Byleen-Higley (Lawrence High School, Lawrence, KS), Adrienne Juett
(KU Physics Dept., currently at MIT),
and Josh Meyers (KU Physics Dept.),
who performed invaluable 
software assistance and antenna and amplifier calibration.
Alexey Provorov and Igor Zheleznykh (Moscow Institute of Nuclear
Research) constructed 
the TEM horn antennas currently used in the surface-noise
veto.
We also thank the winterovers who staffed the
experiment during the last two years at the
South Pole (Xinhua Bai, Alan Baker, Mike Boyce,
Marc Hellwig, Steffen Richter, and Darryn Schneider.)
Ryan Dyer helped in deployment during
the 1999-2000 campaign.
We are indebted to
Dan DePardo and Dilip Tammana for their help in setting up and
operating the KU Antenna Testing Range.
\end{acknowledgements}

\vspace{2cm}
\centerline{\bf{\it APPENDIX 1: Check of Antenna Impedance}}
We can check the internal consistency of our 
${\vec Z_{antenna}}$
measurements by constructing an appropriate Green function
to give the antenna response to a given input. Impedance is,
in effect, a Green function - namely, an analytic function in the
complex $\omega$ plane, reflecting the constraint of causality in
the time domain.

{\it The Zeroes and Poles Expansion} 
of the impedance of a network implements the
essential analytic features.
It states: The impedance of a
network containing any number of $L, \,R,\, C$ elements, including
antennas, is a rational function of $\omega$, which can always be
factored into simple zeroes and poles\cite{ShelkunoffFriis}: 
\ba Z(\omega) = \frac{ ( \omega
-\omega_0) ( \omega -\omega_1) ( \omega -\omega_2) ..  } {i \omega (
\omega -\omega'_0) ( \omega -\omega'_1) ( \omega -\omega'_2) ...  }
\label{impedanceexpansion} .\ea 

The expansion follows from the definition
$Z(\omega)=V(\omega)/I(\omega)$ and the expansion of
$V(\omega)$ and $I(\omega)$ into polynomial ratios,
which can be factored.
The pole at $\omega=0$ is just the static capacitance of the device. 
The poles and zeroes are symmetric about the
imaginary axis and all lie in the lower $\omega$ half-plane, consistent
with causality.

Since all
the poles and zeroes are simple and isolated, an equivalent ansatz is
a sum of familiar ``Breit-Wigner'' resonances, \ba i Z(\omega) =
\sum_{J}\frac{-i \alpha_{J}\omega \omega_J}{ \omega^{2}
-\omega_J^{2}+i \Gamma_{J}\omega }\label{breitwig} .\ea Here $\omega_J$
is the value of a resonant antenna frequency, and $\Gamma_{J}$ is the
width; the constant $ \alpha_{J}$ measures the strength.

\subsubsection*{Check of impedance ansatz with data}

A Hewlett-Packard network analyzer HP8712C measured real
and imaginary reflection
coefficients $\Gamma_{r}, \, \Gamma_{i}$ for
antennas in the lab.  Tests were made
after calibrating out cable, connector and lead effects.  The
reflection coefficients are related to the complex impedance
$Z(\omega)$ by \ba Re(Z(\omega))=
{1-\Gamma_r^2-\Gamma_i^2\over(1-\Gamma_r)^2+\Gamma_i^2} ~{\rm and}~
Im(Z(\omega))={2\Gamma_i\over(1-\Gamma_r)^2+\Gamma_i^2} .\ea  
Variations from antenna to antenna were observed to be small,
typically 10\% or less and considered smaller than other
variables in the calibration chain.

\begin{figure}[htpb]
\begin{picture}(200,250)
\includegraphics{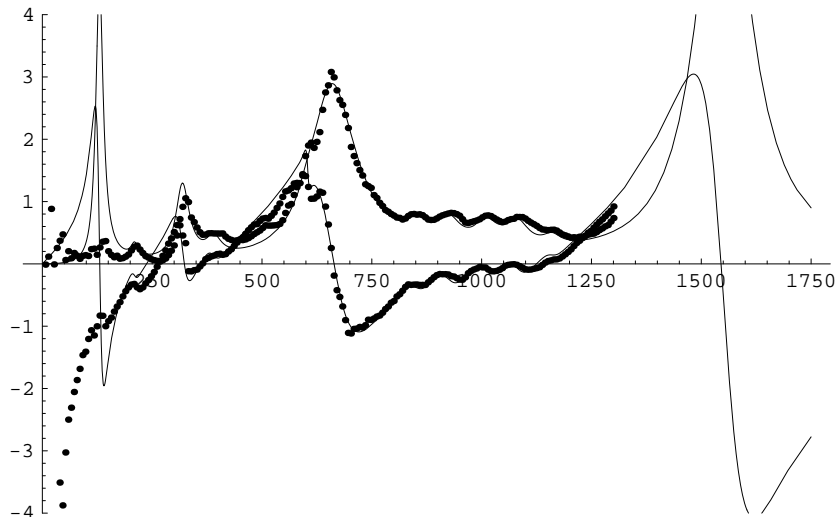}
\end{picture} 
\caption {Real (positive only) and
imaginary (both positive and negative values)
antenna impedance 
data obtained from network analyzer measurements (dots), as a function
of frequency (MHz). The
vertical scale is linear and in absolute units of 50 ohms. 
After performing a multiple Breit-Wigner fit to the real
part of the impedance function $Re(Z(\omega))$, 
the imaginary part of the impedance is predicted
directly from the real part by the dispersion relation.  
Agreement between the calculated and the measured values of
$Im(Z)$ is satisfactory.}
\label{fig:impedanceplot}
\end{figure}

Figure \ref{fig:impedanceplot} shows the real and imaginary parts of
the antenna impedance measurements over the approximate range $50MHz
-5GHz$.  This range substantially exceeds the range over which the
antenna impedance needs to be known, due to cutoffs at the low and
high frequency ends from the high-pass filter and cable losses,
respectively.  These effects are well-characterized and discussed
elsewhere.  The Figure shows real and imaginary parts as ``bumps'' and
wiggle-excursions consistent with Breit-Wigner physics.

We tested our understanding of the antennas by fitting the real part
(only) of the impedance to a sum of Breit Wigner functions.  This is
straightforward because the resonant frequency is located near the
bump maximum and the width can be estimated from the graphs.  Fitting
one-half of the data is the same procedure as fitting a {\it
dispersion relation}, namely \ba iZ(\omega) = \frac{1}{2 \pi i} 
\int_{-\infty}^{+\infty} d
\omega'\, \frac{ Im( iZ(\omega') )}{ \omega-\omega' }.  \nn \ea
Compared to standard dispersion relations, real and imaginary
conventions are reversed in impedance, with the absorptive part of
impedance (resonant bumps) defined as real-valued.

Once the real part was fit, it was used to {\it predict} the imaginary
part of the impedance (Fig.  \ref{fig:impedanceplot}, gray lines). The
predictions agree with the data and serve as a check of the
entire procedure.

\end{document}

%% file: riceauthor.tex
\author{I. Kravchenko}
\address{Massachusetts Institute of Technology Lab. for
Nuclear Science, Cambridge, MA  02139}
\author{G. M. Frichter}
\address{Florida State University, 
High Energy Physics, Tallahassee FL  32306-4350}
\author{D. Seckel, G. M. Spiczak\footnote{present address:
University of Wisconsin, 
River Falls, WI 54022}}
\address{Bartol Research Institute, U. of Delaware, Newark DE 19716}
\author{S. Seunarine}
\address{Department of Physics and Astronomy,
Private Bag 4800, U. of Canterbury, Christchurch, New Zealand}
\author{C. Allen,
A. Bean,
D. Besson,\footnote{Contact: zedlam@ku.edu, for more information on the
RICE experiment.}
D. J. Box,
R. Buniy,\footnote{Currently at Vanderbilt University, Nashville, TN, 37235}
D. McKay,
L. Perry,
J. Ralston,
S. Razzaque,
D. W. Schmitz\footnote{Currently at Columbia University, NYC, NY  10027}}
\address{University of Kansas Dept. of Physics and Astronomy, Lawrence KS
66045-2151}


%% file: astroph_cal.bbl
\begin{thebibliography}{99}

\bibitem{doublebang}J. Learned and S. Pakvasa, hep-ph/9408296

\bibitem{tomography}A. Nicolaidis (Thessaloniki U.), 
M. Jannane, A. Tarantola (Paris, Curie Univ. VI, Inst.
Phys. Globe), preprint {\bf THES-TP-90-03}, Jul 1990.;
P. Jain, J. P. Ralston and G. Frichter, {\it
Astroparticle Physics} {\bf 12}, 193 (1999).  See also J. P. Ralston,
in {\it 26th International Cosmic Ray Conference} (Salt lake City
1999) edited by D. Kieda and B. L. Dingus (IUPAP 1999).

\bibitem{AMANDA99}E. Andres et al. 
Astropart Phys, 13 (2000) 1.
\bibitem{NESTOR}See {\it
Proc. of 3rd NESTOR Intl. Workshop}, Ed.  L.K. Resvanis, 
Athens, Greece, Univ. Press, 1993
\bibitem{BAIKAL99}V. Balkanov {\it et al.}, astro-ph/0001151,
to appear in the Proceedings of International
    Conference on Non-Accelerator 
New Physics, June 28 - July 3, 1999, Dubna, Russia
\bibitem{Antares00}F. Montanet, astro-ph/0001380,
Talk given at TAUP99, the Sixth International Workshop on Topics in
    Astroparticle and Underground Physics, College de France, Paris, 
France, September
    6-10, 1999 
\bibitem{TWR}Monte Carlo simulations have shown that the implementation
of Transient Waveform Recorders (TRW's), planned by the AMANDA collaboration
for use in 2002, should significantly improve sensitivity to electromagnetic
cascades in the ice.
\bibitem{ICRC-papers}The AMANDA Collaboration, 
``Search for Neutrino Induced Cascades with the AMANDA-B10 Detector''
in Proc. of the 26th Intl. Cosmic Ray Conference,
DESY, Zeuthen, Germany (2001)
\bibitem{Buford}
Price, P. B., astro-ph/951011, (1995), and
Astropart. Phys. {\bf 5}, 43, (1996).

\bibitem{Allan} Allan, H. R., {\it Progress in Elementary
Particles and Cosmic Ray Physics}, North-Holland Publishing
Company, Amsterdam, 1971.
\bibitem{Askaryan} Askaryan, G. A., Zh. Eksp. Teor. Fiz.
{\bf 41},616 (1961) [Soviet Physics JETP {\bf 14}, 441 (1962)].
\bibitem {Jelley} Jelley, J. V. et al, Nu. Cim. {\bf X46}, 649
(1966)
\bibitem{Rosner} Rosner, J. L. and Wilkerson, J. F., EFI-97-10
(1997), hep-ex/9702008; Rosner, J. L., DOE-ER-40561-221 (1995),
hep-ex/9508011. Rosner, J. L., Extensive Air Shower Radio Detection:
Recent Results and Outlook, astro-ph/0101089
invited talk presented by J. Rosner at
    RADHEP-2000 Conference, Nov. 16-18, 2000 (UCLA).
\bibitem{Auger00}M. Ave {\it et al.}, astro-ph/0003011
\bibitem{RADHEP2000}See, e.g., Proc. of the 1st International 
Workshop on
Radio Detection of High-Energy Particles (RADHEP-2000), 
Nov 16-18, 2000 (UCLA).
\bibitem{testbeam}D. Saltzberg {\it et. al.}, Phys. Rev. Lett.
\bibitem{fmr}Frichter, G. M., Ralston J. P., and McKay, D. W.,
Phys. Rev. {\bf D 53}, 1684 (1996).


\bibitem{Bog80} V. V. Bogorodsky and V. P. Gavrilo, {\it Ice: Physical
Properties}, Modern Methods of Glaciology (Lenningrad, 1980); V. V. Bogorodsky,
C. R. Bentley and P. E. Gundmandsen, "Radioglaciology", Reidel Publishing 
(1985).

\bibitem{Har65} C. W. Harrison and C. S. Williams, IEEE Trans.
Anten. Prop., {\bf AP-13}, 236 (1965)


\bibitem{ZHS}Zas, E., Halzen, F., and Stanev, T., Phys. Rev.
{\bf D 45}, 362 (1992).

\bibitem{BigPaper}S. Razzaque {\it et. al.}, in preparation, and
S. Razzaque {\it et. al.}, in the Proc. of the 1st International 
Workshop on
     Radio Detection of High-Energy Particles (RADHEP-2000), 
Nov 16-18, 2000 (UCLA).


\bibitem{ShelkunoffFriis}
S. A. Shellkunoff and H. T. Friis, {\it Antennas: Theory and
Practice}, (Wiley, New York, 1952); and
S. A. Shellkunoff , {\it Advanced Antenna Theory}, (Wiley, New York, 1952).




\end{thebibliography}
